\documentclass[aps,prc,twocolumn,superscriptaddress,preprintnumbers,showpacs,nofootinbib,a4paper,amsmath,amssymb]{revtex4-1}

\usepackage{graphicx}
\usepackage{dcolumn}
\usepackage{bm}
\usepackage{color}%
\usepackage{float}
\usepackage[bookmarks=false, pdfborder={0 0 0}, colorlinks=true, linkcolor=blue, citecolor=blue, filecolor=blue, urlcolor=blue]{hyperref}

\graphicspath{{.}{../v3/}}
\newcommand{\hzdr}{\affiliation{Helmholtz-Zentrum Dresden-Rossendorf (HZDR), Bautzner Landstr. 400, D-01328 Dresden, Germany}}
\newcommand{\tu}{\affiliation{Technische Universit\"at Dresden, Institut f\"ur Kern- und Teilchenphysik, D-01062 Dresden, Germany}}
\newcommand{\infnpd}{\affiliation{Istituto Nazionale di Fisica Nucleare (INFN), Sezione di Padova, I-35131 Padova, Italy}}
\newcommand{\unige}{\affiliation{Dipartimento di Fisica, Universit\`a degli Studi di Genova, and Istituto Nazionale di Fisica Nucleare (INFN), Sezione di Genova, Genova, Italy}}
\newcommand{\unipd}{\affiliation{Dipartimento di Fisica e Astronomia, Universit\`a degli Studi di Padova, Padova, Italy}}
\newcommand{\atomki}{\affiliation{Institute for Nuclear Research (MTA ATOMKI), PO Box 51, HU-4001 Debrecen, Hungary}}
\newcommand{\supa}{\affiliation{SUPA, School of Physics and Astronomy, University of Edinburgh, EH9 3FD Edinburgh, United Kingdom}}
\newcommand{\lngs}{\affiliation{Laboratori Nazionali del Gran Sasso (LNGS), 67100 Assergi (AQ), Italy}}
\newcommand{\gssi}{\affiliation{Gran Sasso Science Institute, 67100 L’Aquila, Italy}}
\newcommand{\unina}{\affiliation{Universit\`a degli Studi di Napoli Federico II and INFN, Sezione di Napoli, 80126 Napoli, Italy}}
\newcommand{\unito}{\affiliation{Universit\`a degli Studi di Torino and INFN, Sezione di Torino, Via P. Giuria 1, 10125 Torino, Italy}}
\newcommand{\unimi}{\affiliation{Universit\`a degli Studi di Milano and INFN, Sezione di Milano, Via G. Celoria 16, 20133 Milano, Italy}}
\newcommand{\infnrm}{\affiliation{INFN, Sezione di Roma La Sapienza, Piazzale A. Moro 2, 00185 Roma, Italy}}
\newcommand{\uniba}{\affiliation{Universit\`a degli Studi di Bari and INFN, Sezione di Bari, 70125 Bari, Italy}}
\newcommand{\obste}{\affiliation{INAF Osservatorio Astronomico di Teramo, Teramo, Italy}}

\begin{document}
\title{Direct measurement of low-energy $^{22}$Ne(p,$\gamma$)$^{23}$Na resonances}

\author{R.~Depalo}\unipd\infnpd
\author{F.~Cavanna}\unige
\author{M.~Aliotta}\supa
\author{M.~Anders}\hzdr\tu
\author{D.~Bemmerer}\email[e-mail address:~]{d.bemmerer@hzdr.de}\hzdr
\author{A.~Best}\lngs
\author{A.~Boeltzig}\gssi
\author{C.~Broggini}\infnpd
\author{C.G.~Bruno}\supa
\author{A.~Caciolli}\unipd\infnpd
\author{G. F.~Ciani}\gssi
\author{P.~Corvisiero}\unige
\author{T.~Davinson}\supa
\author{A.~Di~Leva}\unina
\author{Z.~Elekes}\atomki
\author{F.~Ferraro}\unige
\author{A.~Formicola}\lngs
\author{Zs.~F\"ul\"op}\atomki
\author{G.~Gervino}\unito
\author{A.~Guglielmetti}\unimi
\author{C.~Gustavino}\infnrm
\author{Gy.~Gy\"urky}\atomki
\author{G.~Imbriani}\unina
\author{M.~Junker}\lngs
\author{R.~Menegazzo}\infnpd
\author{V.~Mossa}\uniba
\author{F. R.~Pantaleo}\uniba
\author{D. ~Piatti} \unipd\infnpd
\author{P.~Prati}\unige
\author{O.~Straniero}\obste\lngs
\author{T.~Sz\"ucs}\atomki
\author{M.~P.~Tak\'acs}\hzdr\tu
\author{D.~Trezzi}\unimi
\collaboration{The LUNA Collaboration}\noaffiliation		

\date{\today}

\begin{abstract}
\textbf{Background:} The $^{22}$Ne(p,$\gamma$)$^{23}$Na reaction is the most uncertain process in the neon-sodium cycle of hydrogen burning. At temperatures relevant for nucleosynthesis in asymptotic giant branch stars and classical novae, its uncertainty is mainly due to a large number of predicted but hitherto unobserved resonances at low energy.

\textbf{Purpose:} A new direct study of low energy $^{22}$Ne(p,$\gamma$)$^{23}$Na resonances has been performed at the Laboratory for Underground Nuclear Astrophysics (LUNA), in the Gran Sasso National Laboratory, Italy.

\textbf{Method:} The proton capture on $^{22}$Ne was investigated in direct kinematics, delivering an intense proton beam to a $^{22}$Ne gas target. $\gamma$ rays were detected with two high-purity germanium detectors enclosed in a copper and lead shielding suppressing environmental radioactivity.

\textbf{Results:} Three resonances at 156.2 keV ($\omega\gamma$ = (1.48\,$\pm$\,0.10)\,$\cdot$\,10$^{-7}$ eV), 189.5 keV ($\omega\gamma$ = (1.87\,$\pm$\,0.06)\,$\cdot$\,10$^{-6}$ eV) and 259.7 keV ($\omega\gamma$ = (6.89\,$\pm$\,0.16)\,$\cdot$\,10$^{-6}$ eV) proton beam energy, respectively, have been observed for the first time. For the levels at $E_{\rm x}$ = 8943.5, 8975.3, and 9042.4 keV excitation energy corresponding to the new resonances,  the $\gamma$-decay branching ratios have been precisely measured. Three additional, tentative resonances at 71, 105 and 215 keV proton beam energy, respectively, were not observed here. For the strengths of these resonances, experimental upper limits have been derived  that are significantly more stringent than the upper limits reported in the literature. 

\textbf{Conclusions:} Based on the present experimental data and also previous literature data, an updated thermonuclear reaction rate is provided in tabular and parametric form. The new reaction rate is significantly higher than previous evaluations at temperatures of 0.08-0.3 GK. 

\end{abstract}


\maketitle

\section{\label{sec:intro} Introduction}
The neon-sodium cycle of hydrogen burning (NeNa cycle) converts hydrogen to helium through a chain of proton induced reactions involving neon and sodium isotopes. The NeNa cycle contributes negligibly to the energy budget, but it is of great importance for stellar nucleosynthesis, because it affects the abundances of the elements between $^{20}$Ne and $^{27}$Al \cite{Marion57-ApJ}.

Predicting the abundances of NeNa cycle elements in different astrophysical objects has become highly topical since the discovery of the anticorrelation between sodium and oxygen abundances in red giant stars of globular clusters \cite[a review]{Gratton12-AAR}. In principle, material showing such a Na-O anticorrelation may be produced by hydrogen burning when the temperature is so high that not only the CNO cycle, but also the NeNa cycle is activated \cite{Prantzos07-AA}. In order to explain how such material can be brought to the stellar surface where it can be detected, mainly one scenario is invoked: Pollution of the interstellar medium with the ashes of hydrogen burning from a previous generation of stars. Several candidates have been proposed for such a pollution: intermediate-mass Asymptotic Giant Branch (AGB) stars or super AGB stars \cite{Ventura13-MNRAS, Ventura12-ApJ, Doherty14-MNRAS}, fast rotating massive stars \cite{Decressin07-AA}, supermassive stars \cite{Denissenkov15-MNRAS}, massive stars in close binary systems \cite{deMink09-AA}, stellar collisions \cite{Sills10-MNRAS}, and classical novae \cite{Maccarone12-MNRAS}. A detailed understanding of the nuclear physics of the NeNa cycle is needed in order to sharpen quantitative predictions for the nucleosynthetic output of these objects, which, in turn, are necessary to evaluate their respective impact on the Na-O anticorrelation \cite{Denissenkov15-MNRAS}.

\begin{figure}[tb]
\includegraphics[width=7cm]{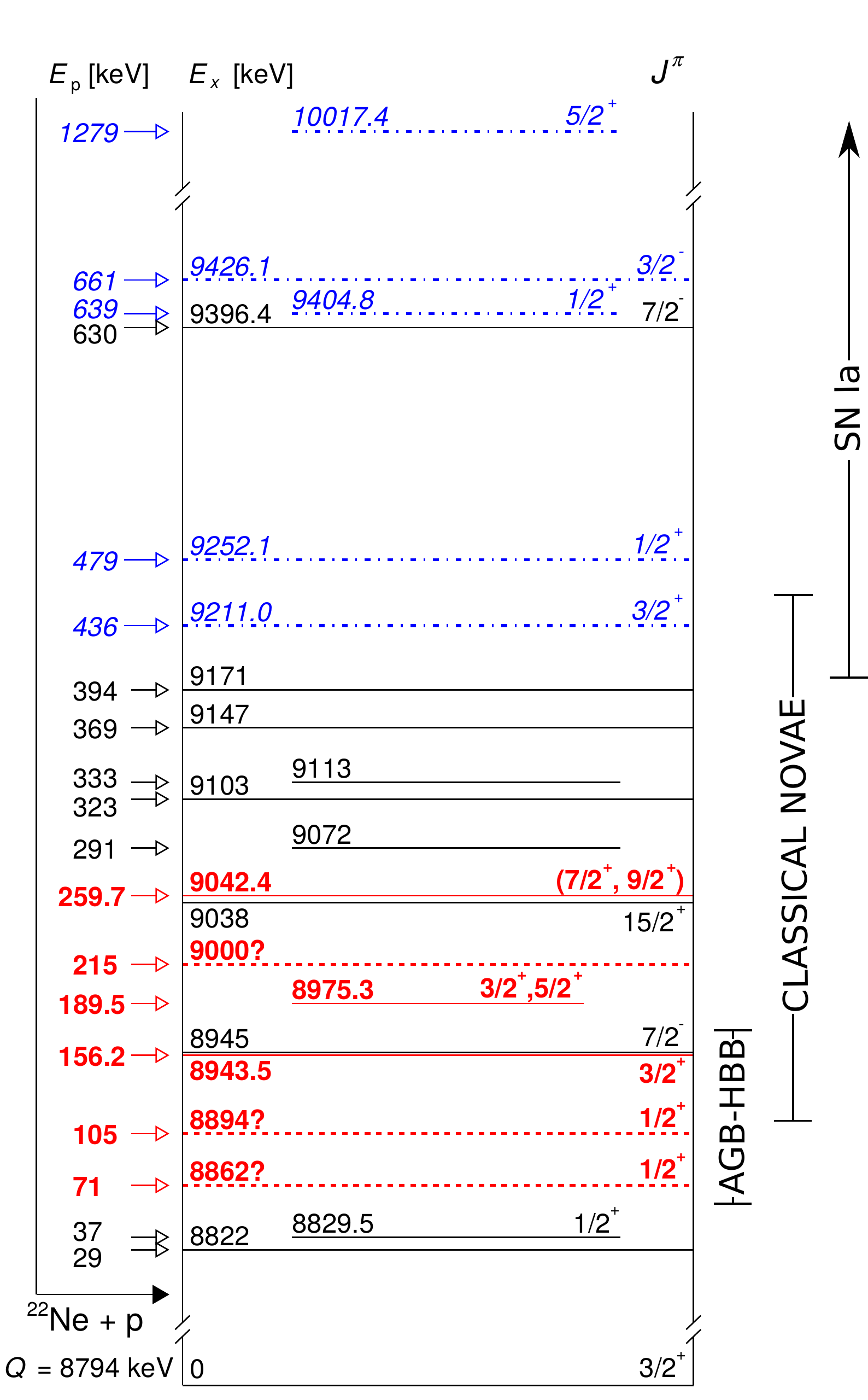}
\caption{\label{fig:ls}(Color online) Partial level scheme of $^{23}$Na, level energies taken from \cite{Firestone07-NDS23,Jenkins13-PRC,Cavanna15-PRL}. The $Q$-value and the resonance energies in the laboratory system of the $^{22}$Ne(p,$\gamma$)$^{23}$Na reaction are also shown. The resonances highlighted in red have been studied at LUNA \cite[and present work]{Cavanna15-PRL}. The resonances shown in blue have recently been investigated at HZDR \cite{Depalo15-PRC}. On the right side of the plot, relevant energy ranges are given for three astrophysical scenarios.}
\end{figure}

The $^{22}$Ne(p,$\gamma$)$^{23}$Na reaction is the most uncertain reaction of the cycle \cite{NACRE99-NPA}. The uncertainty is due to a large number of predicted but hitherto unobserved resonances at proton energies\footnote{Here, $E_{p}$ refers to the proton beam energy in the laboratory system, $E$ to the center-of-mass energy, and $E_{\rm x}$ to the excitation energy in the $^{23}$Na nucleus.} below 400 keV \cite{NACRE99-NPA, Iliadis10-NPA841_31,Iliadis10-NPA841_251,Sallaska13-ApJSS}, and to a number of observed but poorly constrained resonances at 400-1300\,keV \cite{Depalo15-PRC}, see Fig. \ref{fig:ls} for a partial level scheme of $^{23}$Na including the astrophysically relevant resonances.

The resonances above 400 keV have been studied several times before, most recently at the 3\,MV Tandetron of Helmholtz-Zentrum Dresden-Rossendorf (HZDR) \cite[and references therein]{Depalo15-PRC}. For the resonances below 400 keV, however, data are scarce. The  $^{23}$Na  states between $E_{\rm x}$ = 8822-9171 keV have been studied with indirect techniques \cite{Powers71-PRC, Hale01-PRC, Jenkins13-PRC}, but in many cases the spin and parity assignments to those levels are uncertain and their $\gamma$ decay modes are poorly known, if at all. Direct \cite{Goerres82-NPA} and indirect \cite{Powers71-PRC, Hale01-PRC} experiments have been performed to derive the strength of $^{22}$Ne(p,$\gamma$)$^{23}$Na resonances below 400 keV, but both approaches only provided upper limits (tab. \ref{tab:wg_TRR}). Moreover, three states at $E_{\rm x}$ = 8862, 8894 and 9000 keV have been reported as tentative in an indirect measurement \cite{Powers71-PRC}.

The upper limits for the unobserved resonances are treated differently in the literature. In the NACRE compilation \cite{NACRE99-NPA}, usually 1/10 of the upper limit is adopted as value, with an error bar from zero to the full upper limit. In the Iliadis/STARLIB compilation \cite{Iliadis10-NPA841_31,Iliadis10-NPA841_251,Sallaska13-ApJSS}, several unobserved resonances were completely excluded from consideration, therefore also the uncertainty stemming from these cases was left out \cite{Iliadis10-NPA841_251}. For the other resonances, Iliadis/STARLIB adopted a Monte Carlo sampling technique. As a result of the different treatment of the unobserved resonances between NACRE, on the one hand, and Iliadis/STARLIB, on the other hand, the difference between the adopted rates in the two compilations is as high as a factor of 700 at a temperature of 0.08 GK \cite{NACRE99-NPA,Sallaska13-ApJSS}. Sensitivity studies focusing on  nucleosynthesis in AGB stars experiencing Hot Bottom Burning (HBB) \cite{Izzard07-AA} and on classical novae explosions \cite{Iliadis02-ApJSS} show how the uncertainty in the $^{22}$Ne(p,$\gamma$)$^{23}$Na rate affects the abundances of the nuclides between $^{20}$Ne and $^{27}$Al, leading to abundance variations as high as two orders of magnitude.

In order to clarify the situation of the $^{22}$Ne(p,$\gamma$)$^{23}$Na resonances below 400 keV, a new direct experiment has been performed deep underground at the Laboratory for Underground Nuclear Astrophysics (LUNA) \cite{Broggini10-ARNPS}.

The present data on the $E_{p}$ = 156.2, 189.5 and 259.7 keV resonances has been published in abbreviated form \cite{Cavanna15-PRL}. Here, full details of the experiment and data analysis are given, together with the decay branching ratios of the newly observed resonances. The thermonuclear reaction rate is given both in tabular and parameterized forms for use in stellar evolution codes. 

\begin{figure*}[t!!]
\includegraphics[trim=3cm 4cm 3cm 4cm,width=11cm]{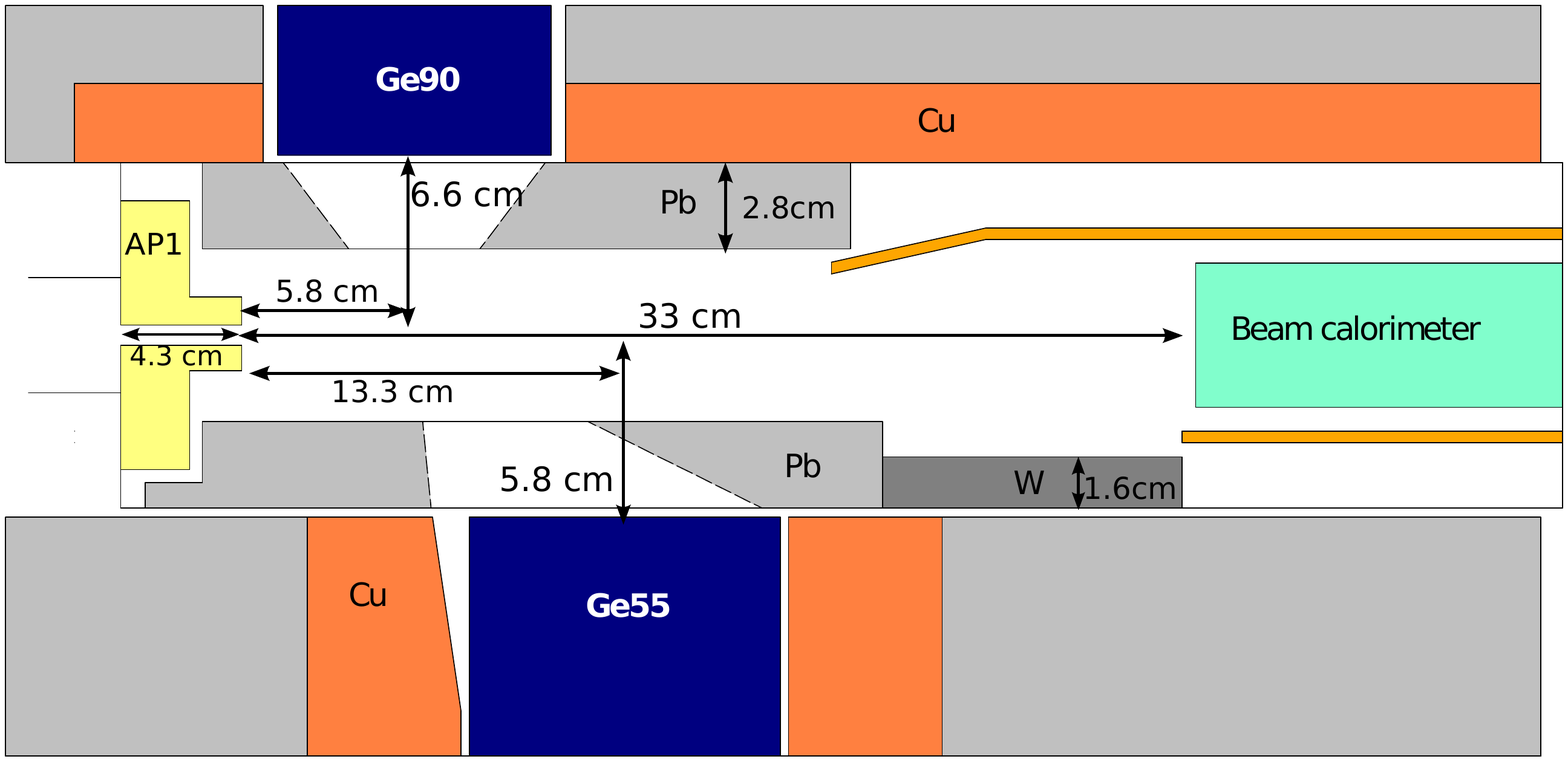}
\caption{\label{fig:chamber}(Color online) Sketch of the experimental setup. The proton beam enters the target chamber from the left hand side of the picture and stops on the beam calorimeter. The collimation system for the germanium detectors is shown in detail.}
\end{figure*}

The present work is organized as follows. The experimental method and the setup are described in sec. \ref{sec:experiment}. Section \ref{sec:results} discusses the data analysis technique. The results obtained for each of the resonances investigated are shown in secs. \ref{subsec:156}-\ref{subsec:ul}. The astrophysical reaction rate and a brief discussion are given in sec. \ref{sec:Astro}. 

\section{\label{sec:experiment} Experiment}
The experiment has been conducted at the LUNA 400 kV accelerator \cite{Formicola03-NIMA}, deep underground at the Gran Sasso National Laboratories (LNGS), operated by the Italian National Institute for Nuclear Physics (INFN). Owing to its deep underground location, LUNA affords a uniquely low background \cite{Costantini09-RPP, Broggini10-ARNPS}, allowing to investigate proton capture on $^{22}$Ne with unprecedented sensitivity \cite{Cavanna14-EPJA}. 

\subsection{Ion beam and target chamber}
A proton beam with a typical intensity of 200 $\mu$A was delivered to a differentially pumped, windowless gas target \cite[for details]{Cavanna14-EPJA}. The target chamber was filled with 1.5 mbar of neon gas, enriched to 99.9$\%$ in $^{22}$Ne. A recirculation and purification system collected the gas exhaust from the pumping stages, removed possible contaminants through a chemical getter and finally recycled the gas to the target chamber, keeping the purity of $^{22}$Ne constant throughout the measurements. 

The beam entered the target chamber through a water cooled copper collimator (denoted AP1 in fig. \ref{fig:chamber}) of 4.3 cm length and 0.7 cm diameter and was stopped on the hot side of a beam calorimeter, used to measure the beam intensity with 1$\%$ precision \cite{Casella02-NIMA, Cavanna14-EPJA}.

The gas density as a function of the position inside the chamber and the beam heating correction have been studied in details \cite{Cavanna14-EPJA}, giving the areal number density of the target atoms with an overall uncertainty of 1.3$\%$. 
The possible presence of an air contamination in the gas was checked periodically using the intense $^{14}$N(p,$\gamma$)$^{15}$O resonance at $E_{p}$ = 278 keV. The nitrogen level in the gas was always found to be below 0.1$\%$.

\subsection{$\gamma$-ray detection system}

\begin{figure*}[ht]
\includegraphics[width=\columnwidth]{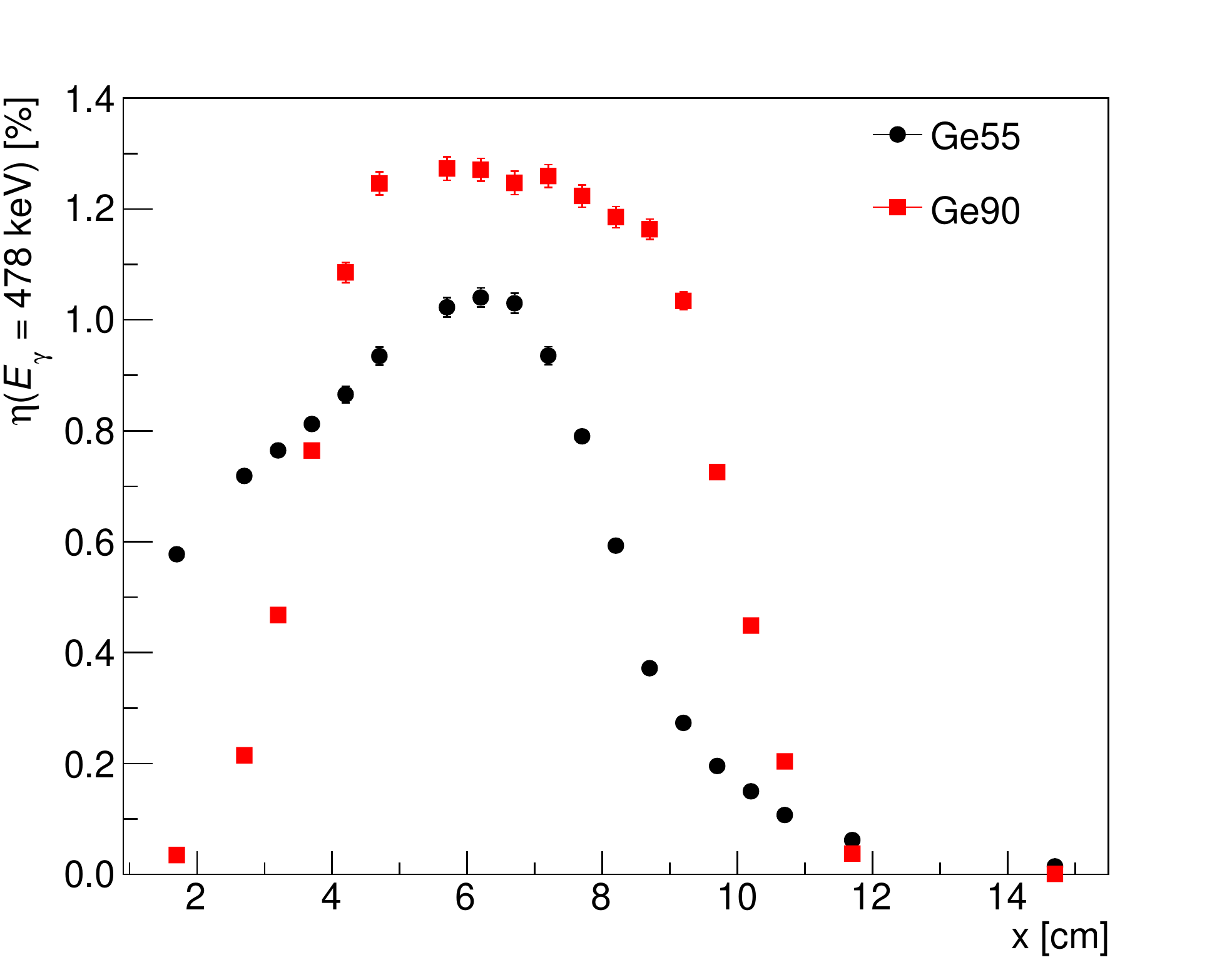}
\includegraphics[width=\columnwidth]{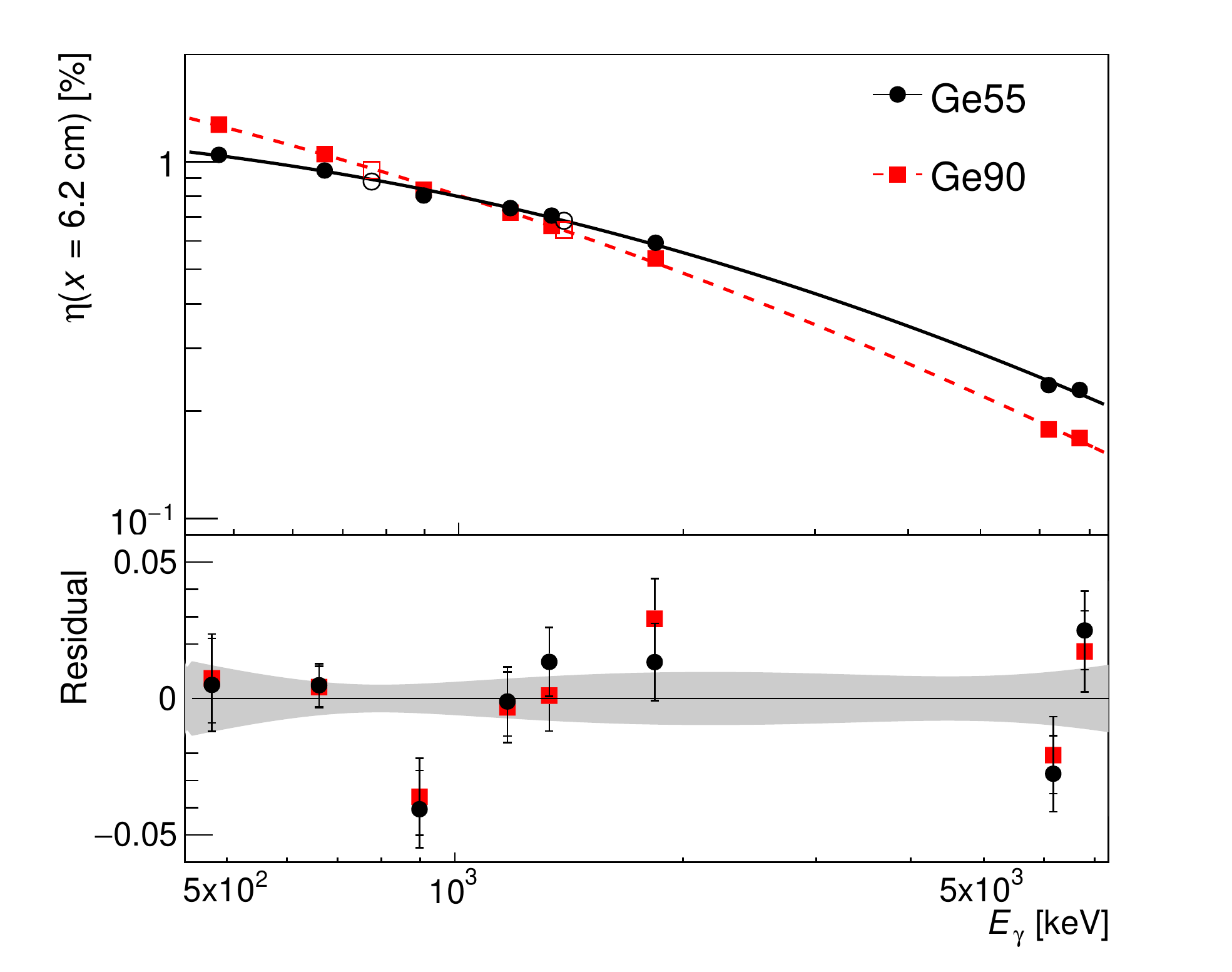}
\caption{\label{fig:efficiency}(Color online) Left: Ge90 (red squares) and Ge55 (black dots) detection efficiency as a function of the distance from the AP1 collimator measured with a $^{7}$Be source. Right: Ge90 (red squares) and Ge55 (black dots) efficiency as a function of the $\gamma$-ray energy measured at $x$ = 6.2\,cm distance from the AP1 collimator. The fit residuals and the 1$\sigma$ uncertainty on the fit function are shown in the lower part of the right panel. Open symbols represent the 765 and 1384 keV gammas from the $^{14}$N(p,$\gamma$)$^{15}$O reaction, which were used to normalize the detection efficiency at 6 MeV (see text for details). }
\end{figure*}

Emitted $\gamma$ rays were detected by two high-purity Germanium (HPGe) detectors (fig. \ref{fig:chamber}): One, of 90$\%$ relative efficiency \cite{Gilmore08-Book},
was mounted on top of the target chamber and was collimated at 90$^{\circ}$ (Ge90). The other, of 135$\%$ relative efficiency, was mounted below the target chamber and collimated at 55$^{\circ}$ (Ge55). Both detectors were enclosed in a copper and lead shield of 15-25\,cm thickness, absorbing $\gamma$ rays from the decay of environmental radioactive isotopes \cite{Cavanna14-EPJA}.

Two independent data acquisition systems were used in parallel: one with a 100 MHz, 14-bit CAEN N1728B digital ADC providing list mode data and the other with a standard analog signal amplification chain and 16k channel histogramming Ortec 919E ADC unit. 

The detection efficiency for $\gamma$-ray energies between 478 and 1836 keV was measured with point-like $^{137}$Cs, $^{60}$Co and $^{88}$Y activity standards from Physikalisch-Technische Bundesanstalt, calibrated with $<1\%$ error, and one $^{7}$Be radioactive source produced at MTA ATOMKI (Hungary) and calibrated with  $1.5\%$ uncertainty. By moving each of the radioactive sources, in turn, along the beam axis, the peak detection efficiency was measured as a function of the distance from the AP1 collimator (see the left panel of fig. \ref{fig:efficiency} for the efficiency profile at $E_\gamma$ = 478\,keV, measured with the $^{7}$Be source). The shape of the profile is dominated by the collimation geometry. For both detectors, the efficiency profile is maximal at a distance $z$ = 6.2 cm from AP1 (fig.\,\ref{fig:efficiency}, left panel). 

The efficiency curve as a function of $\gamma$-ray energy was then extended up to $E_\gamma$ = 6791 keV using the well-known $^{14}$N(p,$\gamma$)$^{15}$O resonance at $E_{p}$ = 278 keV \cite{Formicola04-PLB} ($E_{\rm x}$ = 7556 keV). This resonance de-excites either directly to the ground state or to one of the excited states at 5180, 6172, and 6791 keV. Each of these three states, in turn, decays with 100$\%$ probability directly to the ground state. In the cascade via the 6172 keV (6792 keV) level, a 1384 keV (765 keV) $\gamma$ ray is emitted, lying in the energy range where the efficiency can be determined using radioactive sources. This allows to determine the efficiency for 6 MeV $\gamma$ rays using a relative approach: The efficiency at 1384 keV (765 keV) is derived from the fit of the radioactive sources data, then the efficiency at higher energies was calculated normalizing the 6172 keV (6791 keV) counting rate to the counting rate of the 1384 keV (765 keV) peak. 

Given the short beam-detectors distance, true coincidence summing had to be taken into account in the evaluation of the detection efficiency. The summing-out correction \cite{Gilmore08-Book} depends both on the full-energy peak efficiency and on the total detection efficiency. A Geant4 \cite{Agostinelli03-NIMA} simulation of the setup was used to determine the peak-to-total ratio and to infer the total detection efficiency for $^{60}$Co, $^{88}$Y and $^{14}$N(p,$\gamma$)$^{15}$O $\gamma$ rays. The summing out correction was always between 0.028 and 0.040, depending on $\gamma$-ray energies. A conservative uncertainty of $\pm 20\%$ on the summing-out correction was adopted.

Fig. \ref{fig:efficiency} (right) shows the full-energy peak efficiency as a function of the $\gamma$-ray energy, measured at 6.2 cm from the AP1 collimator. Fits of the efficiency curves with second-order polynomials in the double-logarithmic plane \cite{Knoll10-book} are also shown, together with the fit residuals. The total uncertainty on the detection efficiency determined through the fit functions in fig. \ref{fig:efficiency} is always at the 1-2$\%$ level.

Further details on the experimental setup, and on the data analysis, can be found in thesis works \cite{Cavanna15-PhD,Depalo15-PhD}.

\section{Data analysis and results}
\label{sec:results}

For each resonance studied here, as a first step a run was performed at the beam energy where the maximum yield was expected according to the literature resonance energy, the beam energy loss and the efficiency profile of the detectors. If a signal was observed, a resonance scan was performed, changing the beam energy in 1-2 keV energy steps and measuring the yield of the $E_\gamma$ = 440 keV transition from the first excited state to the ground state of $^{23}$Na. Finally, a long run was performed at the beam energy where the maximal yield was observed, aiming for a statistical uncertainty below 10$\%$ for each detector.

The branching ratio $B_x$ for each transition $x$ was then derived using the following relationship:
\begin{equation}
B_x = \frac{N_x/\eta_x}{\sum_{i}{\left({N_i/\eta_i} \right)}}
\label{eq:branch}
\end{equation}
where $N_i$ is the number of counts observed for each $\gamma$ line and $\eta_i$ its peak detection efficiency. As the $\gamma$-ray spectra were always dominated by the resonance under study, the contribution from possible unobserved transitions was assumed to be negligible in the  branching ratio analysis. Only one detector, Ge55, was used for the branching ratios determination. The second order Legendre polynomial vanishes at 55$^\circ$, minimizing possible angular distribution effects.

\begin{figure*}[t!]
\includegraphics[width=0.32\textwidth]{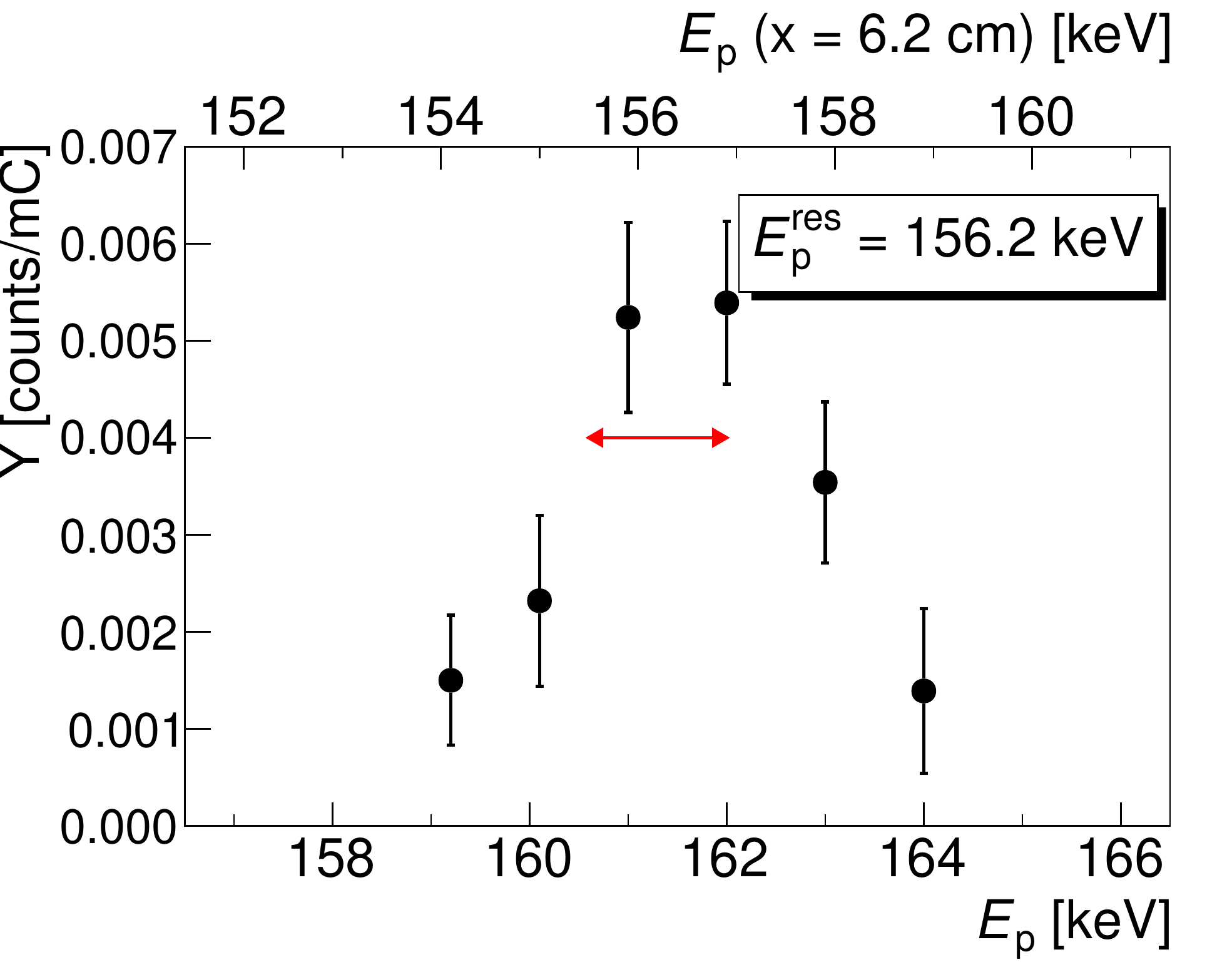}
\includegraphics[width=0.32\textwidth]{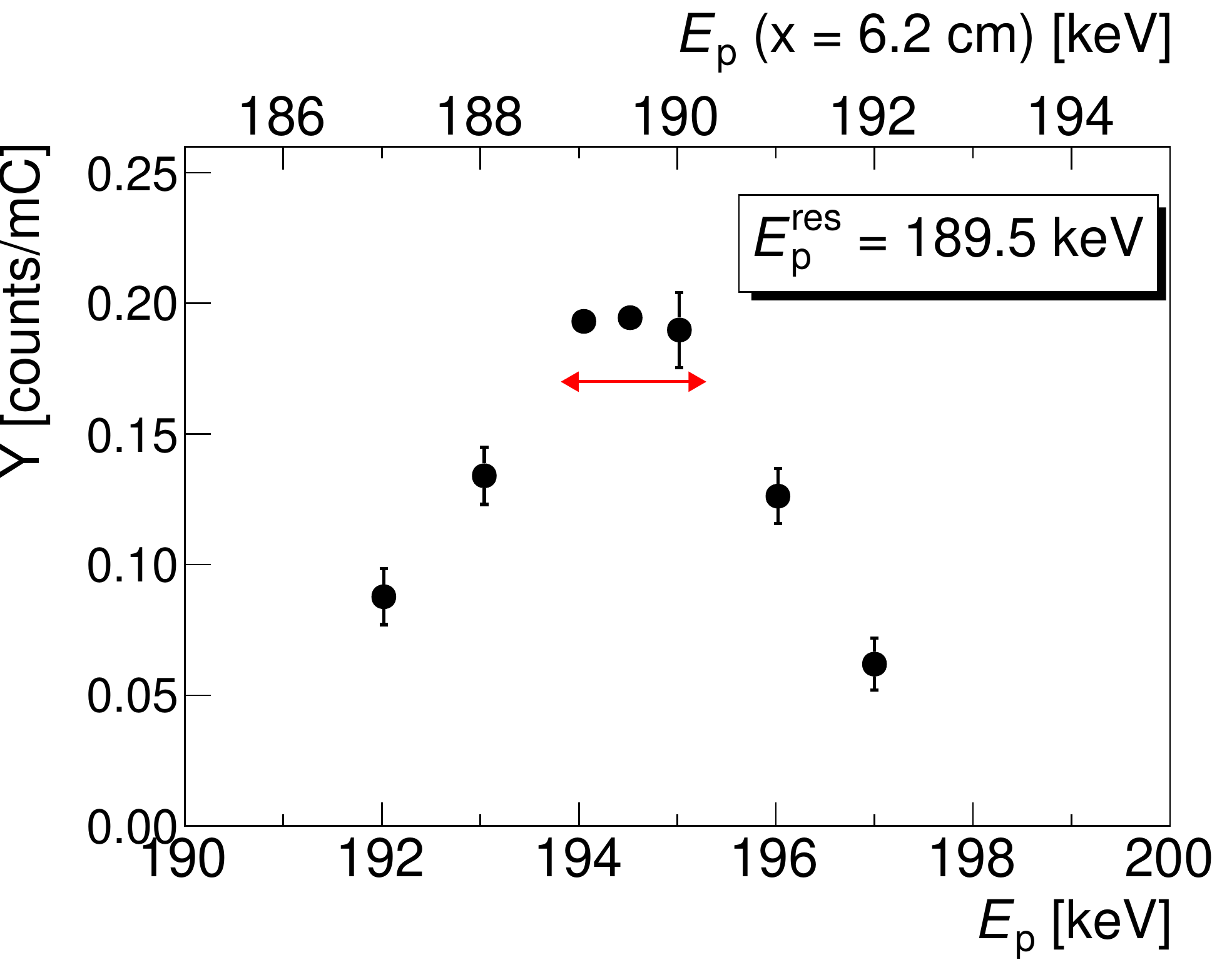}
\includegraphics[width=0.32\textwidth]{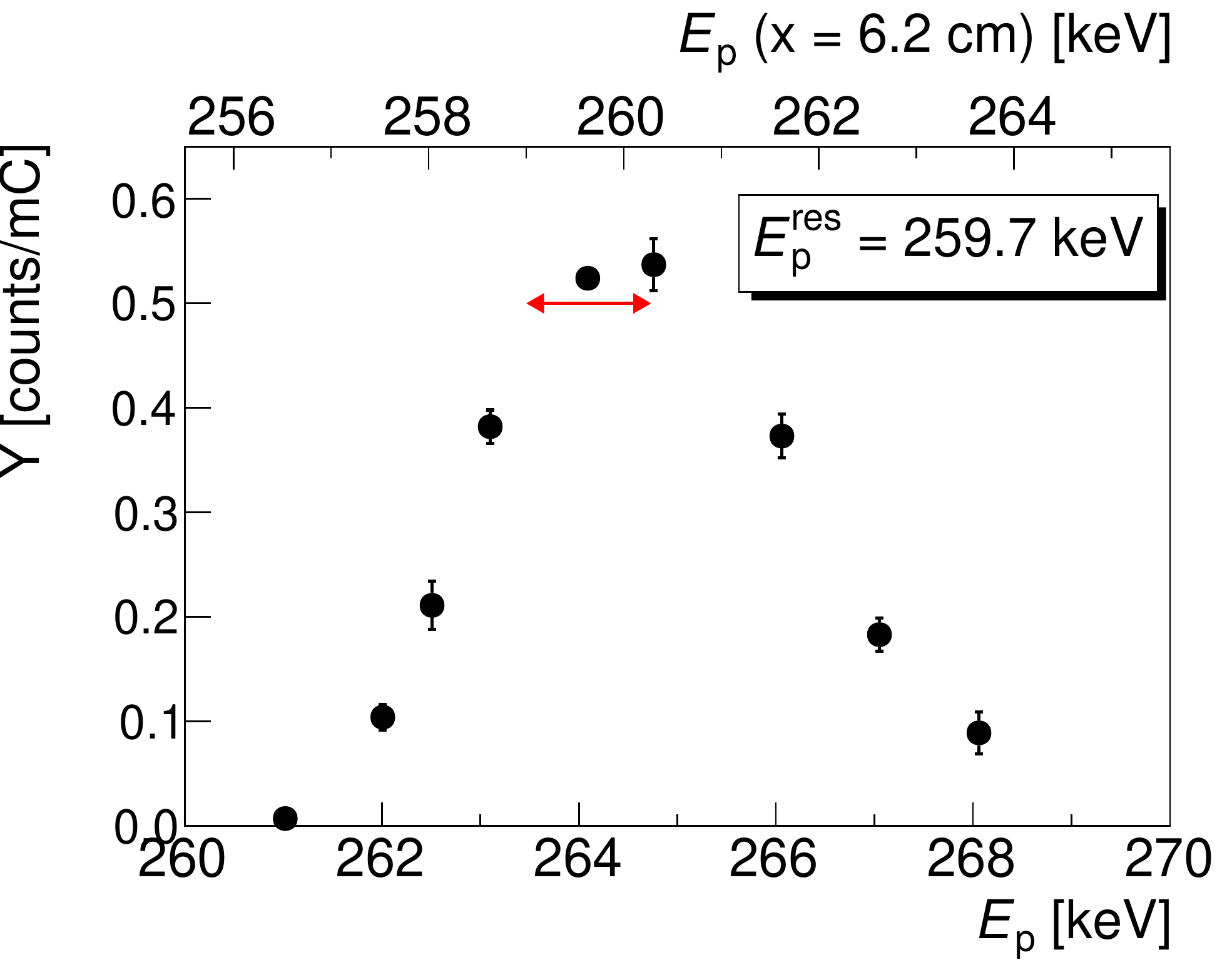}
\caption{\label{fig:scan} Resonance scans for the three newly discovered resonances, measured with Ge55. The bottom x axis shows the initial proton energy. The top axis shows the calculated proton energy at the position where Ge55 has maximum efficiency. The red arrows correspond to the error band assumed for the beam energy at resonance maximum.}
\end{figure*}

In order to determine the energy of the three newly discovered resonances, each resonance scan was compared to the efficiency profile measured with the $^7$Be source. The maxima of the efficiency profile and of the resonance scan are expected to be in the same position along the target chamber, because the two $\gamma$-ray energies are very close by ($E_{\gamma}$\,=\,478\,keV  for $^7$Be, and $E_{\gamma}$\,=\,440\,keV for the resonance scan). The resonance energies were then calculated using the relationship:
\begin{equation}
E_{p}^{\rm res} = E_{\rm p}-(x+x_{\rm coll})\,\frac{dE}{dx}-\Delta E_{\rm pipe}
\end{equation}  
where:
\begin{itemize}
\item $E_{p}$ is the proton energy corresponding to the maximum in the yield profile of the resonance,
\item $x$ is the distance between the place of the efficiency maximum (fig.\,\ref{fig:efficiency}) and the right edge of the AP1 collimator (fig.\,\ref{fig:chamber}),
\item $x_{\rm coll}$ is the effective length of the AP1 collimator (fig.\,\ref{fig:chamber}),
\item $dE/dx$ is the proton energy loss per unit length in neon gas, and
\item $\Delta$$E_{\rm pipe}$ $\approx$ 0.5\,keV is a correction introduced to account for the proton energy loss in the tube that connects the first pumping stage with the target chamber \citep{Cavanna14-EPJA}.
\end{itemize}

The uncertainty of $E_{p}^{\rm res}$ was estimated taking into account the systematic uncertainty on the proton beam energy (0.3\,keV \cite{Formicola03-NIMA}) and on the proton energy loss in neon gas (1.7\%\, \cite{Ziegler10-NIMB}). The overall uncertainty is two to three times smaller than the uncertainty reported in the literature \cite{Iliadis10-NPA841_251, Jenkins13-PRC}. 

The results on the resonance energies are shown in table \ref{tab:res_en}, together with a comparison with previously reported values \cite{Iliadis10-NPA841_251, Jenkins13-PRC}. The values by Jenkins {\it et al.} \cite{Jenkins13-PRC} for the excitation energy $E_{\rm x}$  have been reported with 1\,keV uncertainty. For comparison, the Jenkins $E_{\rm x}$ data are converted to resonance energies $E_{p}^{\rm res}$ using the $Q$-value of the $^{22}$Ne(p,$\gamma$)$^{23}$Na reaction, which is precisely known, $Q$ = (8794.11$\pm$0.02) keV \cite{Audi12-Ame2012}. Also, new recommended excitation energies $E_{\rm x}$ are computed from the present $E_{p}^{\rm res}$ values (table \ref{tab:res_en}).

\begin{table}[tb]
\caption{Resonance energy $E_{p}^{\rm res}$ in keV obtained here (LUNA), compared to the literature values by Iliadis {\it et al.} \cite{Iliadis10-NPA841_251} and by Jenkins {\it et al.} \cite{Jenkins13-PRC}. The values for the excitation energy $E_{\rm x}$ from the present work (LUNA) and from Jenkins {\it et al.} \cite{Jenkins13-PRC} are also shown.}
\begin{center}
\label{tab:res_en}
\begin{tabular}{ c | c | c | c | c }
\hline
$E_{p}^{\rm res}$ LUNA & $E_{p}^{\rm res}$ \cite{Iliadis10-NPA841_251}  &  $E_{p}^{\rm res}$ \cite{Jenkins13-PRC} & $E_{\rm x}$ LUNA & $E_{\rm x}$ \cite{Jenkins13-PRC}\\
\hline
\hline
156.2\,$\pm$\,0.7 & 158\,$\pm$\,2 & 156.8\,$\pm$\,1.0  & 8943.5\,$\pm$\,0.7 & 8944\,$\pm$\,1 \\ 
189.5\,$\pm$\,0.7 & 186\,$\pm$\,2 & 186.0\,$\pm$\,1.0 & 8975.3\,$\pm$\,0.7 & 8972\,$\pm$\,1 \\ 
259.7\,$\pm$\,0.6 & 259.2\,$\pm$\,1.0 & 258.2\,$\pm$\,1.0 & 9042.4\,$\pm$\,0.6 & 9041\,$\pm$\,1  \\ 
\hline          
\end{tabular}
\end{center}
\end{table}%

The present new resonance energies are consistent with the literature, with the exception of the 189.5\,keV resonance, which was expected to be found at 186\,keV \cite{Iliadis10-NPA841_251,Jenkins13-PRC}. The resonance energy enters the reaction rate calculation in an exponential term (eq. \ref{eq:tnrr}). Therefore, at $T_9$ = 0.2 this seemingly small 3.5\,keV shift in resonance energy leads to 18\% reduction in the thermonuclear reaction rate contributed by this resonance, underlining the importance of a correct energy determination. 

For cases where the statistics collected on single $\gamma$-ray transitions was sufficient, the coincidences between the two detectors were studied, in order to confirm that the newly observed transitions are actually related to the decay of the resonance under study. 

\begin{figure*}[t]
\includegraphics[width=\textwidth]{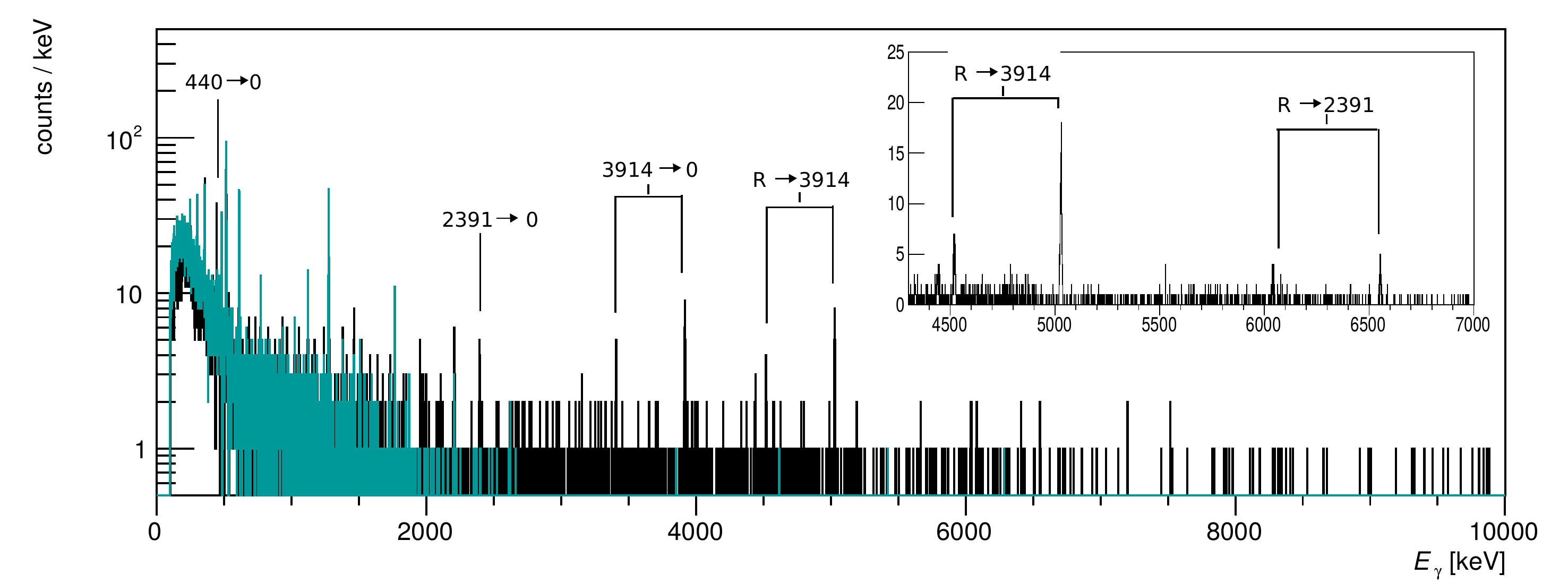}
\caption{\label{fig:156keV_spec} (Color online) $\gamma$-ray spectra measured at the proton energy of maximum yield of the 156.2 keV resonance (black) and off-resonance, i.e. 53 keV below the maximum (turquoise). The insert shows the high-energy part of the spectrum enlarged. }
\end{figure*}

Finally, the resonance strength was deduced from the measured yield, assuming an infinitely thick target, i.e. the proton width $\Gamma_p$ was assumed to be much smaller than 4\,keV (the typical proton beam energy range covered by the HPGe detectors, fig.\,\ref{fig:scan}). Then, the following relationship \cite{Iliadis07-Book} applies:
\begin{equation}
\omega\gamma = \epsilon_R \frac{Y_{max}}{\lambda^2_R/2} \frac{m_t}{m_t+m_p}
\label{eq:wg}
\end{equation} 
where $\epsilon_R$ is the effective stopping power in the laboratory system, $\lambda_R$ is the centre of mass de Broglie wavelength at the resonance energy, and $m_t$ and $m_p$ are the masses of the target and projectile, respectively. The total reaction yield $Y_{\rm max}$ was obtained with the following formula:
\begin{equation}
Y_{\rm max}=\frac{1}{Q}\sum_i\left( \frac{N_i}{\eta_i}\right) 
\label{eq:Ymax}
\end{equation} 
where Q is the total integrated charge, $N_i$ is the number of counts measured for the primary transitions $i$ and $\eta_i$ is the corresponding detection efficiency.  With this approach, the resonance strength is independent of the branching ratios found.

The systematic uncertainty on the resonance strength includes 1.7\% uncertainty on the effective stopping power \cite{Ziegler10-NIMB}, 1\% uncertainty on the beam current determination and an uncertainty between 1\% and 2\% on the detection efficiency.

The tentative resonances at 71, 105 and 215 keV have also been investigated but no $\gamma$ lines from the decay of $^{23}$Na were observed, therefore new upper limits on the resonance strengths were determined.

In the following section, the results obtained for each resonance are discussed.

\subsection{156.2 keV resonance ($E_{\rm x}$ = 8943.5 keV)}
\label{subsec:156}

By strength, the 156.2\,keV resonance is the weakest resonance observed in this experiment. However, it has a significant astrophysical impact, as discussed below in sec.\,\ref{sec:Astro}. The excitation function and a typical spectrum of the resonance are shown in figures \ref{fig:scan} and \ref{fig:156keV_spec}, respectively. Figure \ref{fig:156keV_spec} shows on and off resonance spectra obtained irradiating the target with similar integrated charges. 

$E_p^{\rm res}$ = 156.2\,keV corresponds to 8943.5 keV excitation energy in $^{23}$Na. Jenkins {\it et al.} reported a doublet of two levels at 8944\,keV excitation energy \cite{Jenkins13-PRC}, with tentative spin and parity assignments of 3/2$^+$ and 7/2$^-$, respectively. For the low proton beam energies used here, the 7/2$^-$ level is strongly disfavoured by the angular momentum barrier. Therefore it is assumed that only the 3/2$^+$ level is populated.

Jenkins {\it et al.} report two $\gamma$ rays de-exciting the 3/2$^+$ level, 8944$\rightarrow$2391 and 8944$\rightarrow$3914. The ratio of their respective intensities is 0.7$^{+0.8}_{-0.3}$, in fair agreement with the present, much more precise value of 0.30$\pm$0.05.  The latter value has been calculated from the present absolute branching ratios (table \ref{tab:branching}) for the decay of this level.
The only transition from the de-excitation of the 7/2$^-$ state reported by Jenkins, 8944$\rightarrow$2703, was not detected here, as expected based on the above considerations.  

The single yields measured with each of the two detectors are mutually compatible within statistical errors, therefore no correction was made for the angular distribution, and the total yield was simply taken as the average of the Ge55 and Ge90 yields. 

The final resonance strength is $\omega\gamma = (1.48 \pm 0.10)\times 10^{-7}$ eV. This value is compatible with the upper limit provided by the previous direct experiment \cite{Goerres82-NPA} (see tab. \ref{tab:wg_TRR}). Surprisingly, it is a factor of 16 higher than the upper limit for the resonance strength given by the most recent indirect study, $\omega\gamma$ $\leq$ 0.092$\times$10$^{-7}$ eV \cite{Hale01-PRC}, which is the value adopted by Iliadis/STARLIB \cite{Iliadis10-NPA841_31,Sallaska13-ApJSS}. 

Ref. \cite{Hale01-PRC} assumed the level to have $J^\pi$ = 7/2$^-$, meaning their deduced spectroscopic factor and resonance strength probably refer to the second level of the doublet, not observed here. The existence of a second state with $J^\pi$ = 3/2$^+$ at the same energy was only reported much later \cite{Jenkins13-PRC}. It should be noted that an earlier indirect resonance strength determination, based on a spectroscopic factor from (d,n) stripping reactions and also assuming $J^\pi$ = 7/2$^-$, gave  a strength of $\omega\gamma$ = 6.5$\times$10$^{-7}$ eV \cite{Goerres83-NPA}. The present, direct resonance strength lies between the two indirect numbers \cite{Hale01-PRC,Goerres83-NPA}, which differ by a factor of 70.  

\begin{figure*}[!]
\includegraphics[width=\textwidth]{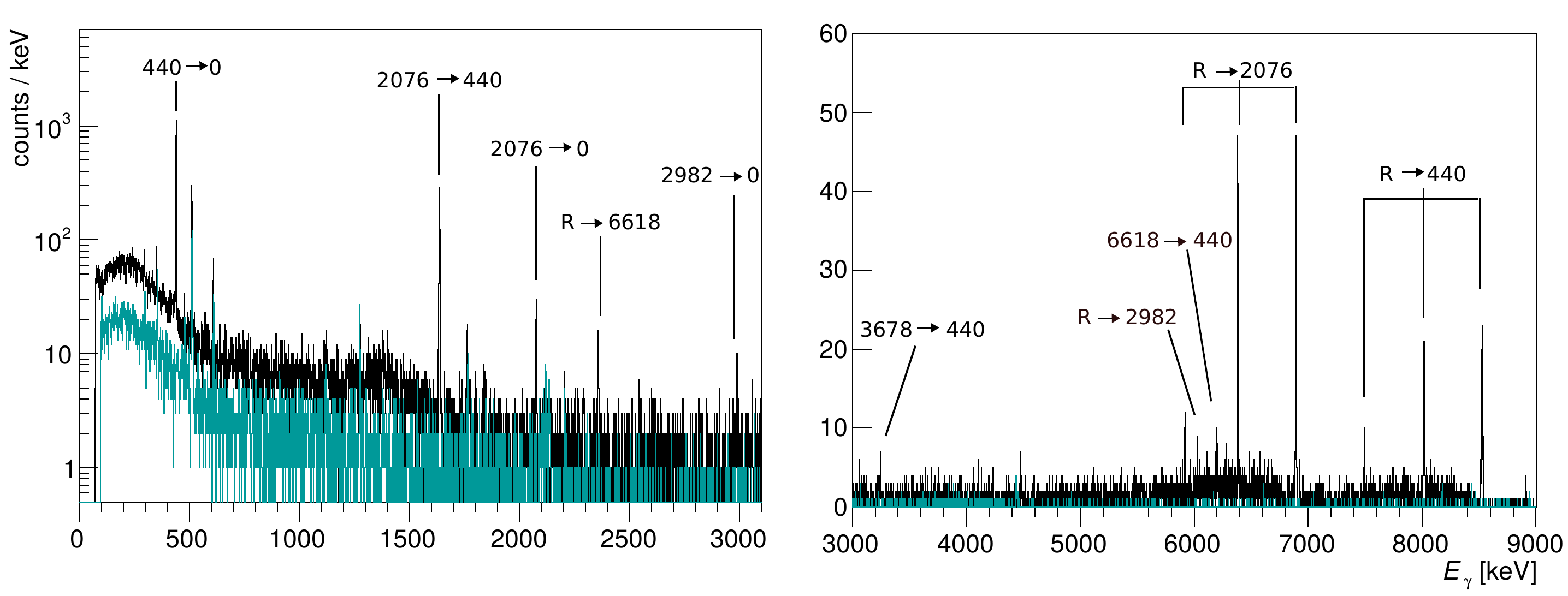}
\includegraphics[width=\textwidth]{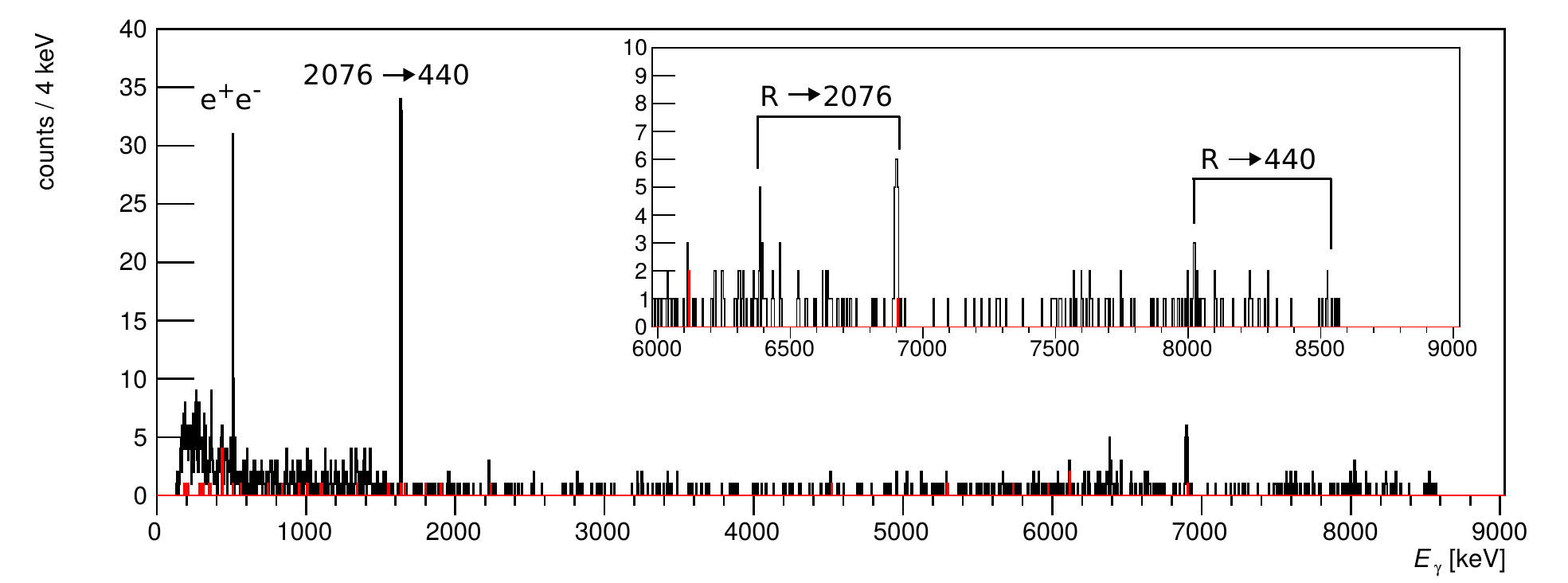}
\caption{\label{fig:189keV_spec} (Color online) Top: $\gamma$-ray spectra measured at the proton energy of maximum yield of the 189.5 keV resonance (black) and 16 keV above the maximum (turquoise). Bottom: coincidence spectrum obtained gating on the 440 keV peak in Ge90 (black) and on a background region adjacent to the 440 keV peak (red). The inset enlarges the high-energy part of the spectrum.}
\end{figure*}

\subsection{189.5 keV resonance ($E_{\rm x}$ = 8975.3 keV)}
\label{subsec:189}

The 189.5\,keV resonance, which corresponds to the 8975.3\,keV excited state in $^{23}$Na, strongly affects the $^{22}$Ne(p,$\gamma$)$^{23}$Na reaction rate for classical novae. The resonance energy is 3.5\,keV higher than previously assumed (fig.\,\ref{fig:scan}). Figure \ref{fig:189keV_spec} shows the comparison between the on- and off-resonance spectra. 

The only literature information on the $\gamma$ decay of the 8975.3\,keV level comes from Ref. \cite{Jenkins13-PRC}, where this level was reported at 8972\,keV. In Ref. \cite{Jenkins13-PRC}, only a transition to the 2982 keV state was observed. In the present work instead, in addition to this transition, five others were also observed, and two of the new branches are each ten times stronger than the only previously reported branch 8975$\rightarrow$2982 (tab.\,\ref{tab:branching}). It can only be speculated that these strong transitions were obscured by some background in Ref. \cite{Jenkins13-PRC}, where many states were populated at the same time, with the entry point at 13\,MeV excitation above the $^{23}$Na ground state.

As an additional cross-check, the most intense among the new transitions was studied by the analysis of $\gamma$-$\gamma$ coincidences between the two detectors: a coincidence gate was applied on the 440 keV transition in Ge90, then the coincident spectrum of Ge55 was analyzed. The coincidence spectrum obtained is shown in fig. \ref{fig:189keV_spec}, compared to a coincidence spectrum obtained gating on a background region next to the 440 keV peak and as wide as the region of interest of the 440 keV peak.

\begin{figure*}[!]
\includegraphics[width=\textwidth]{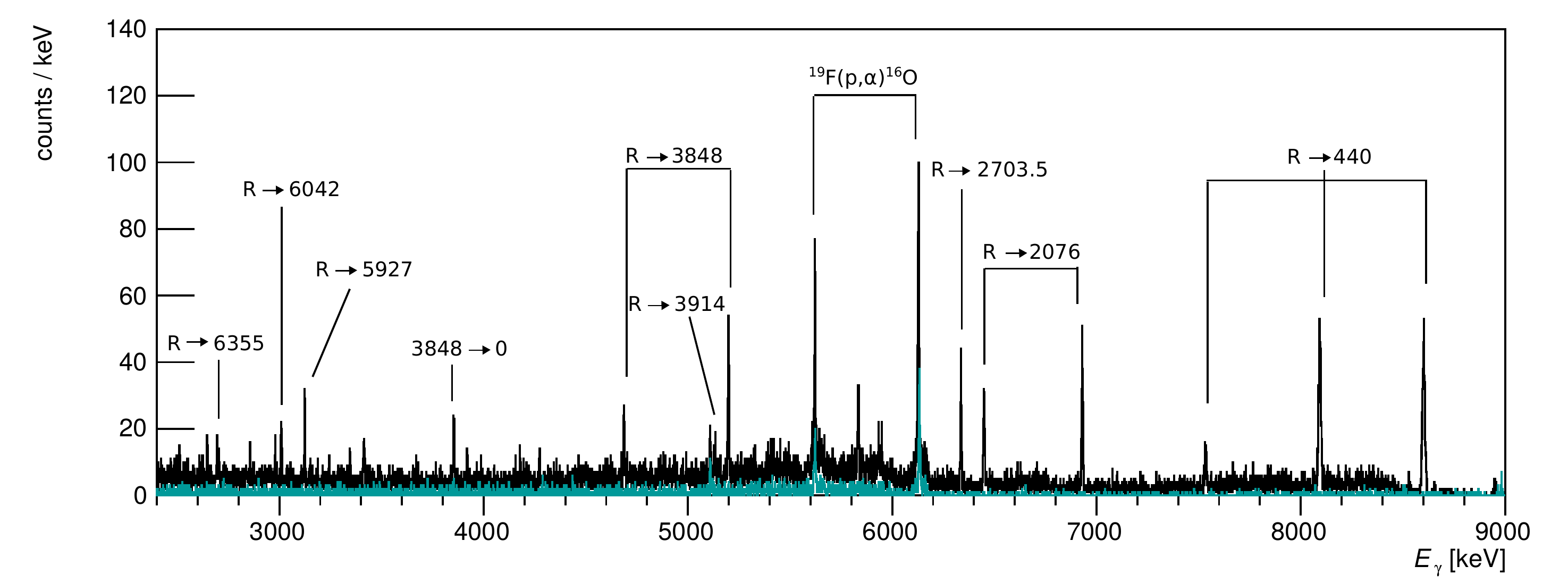}
\includegraphics[width=\textwidth]{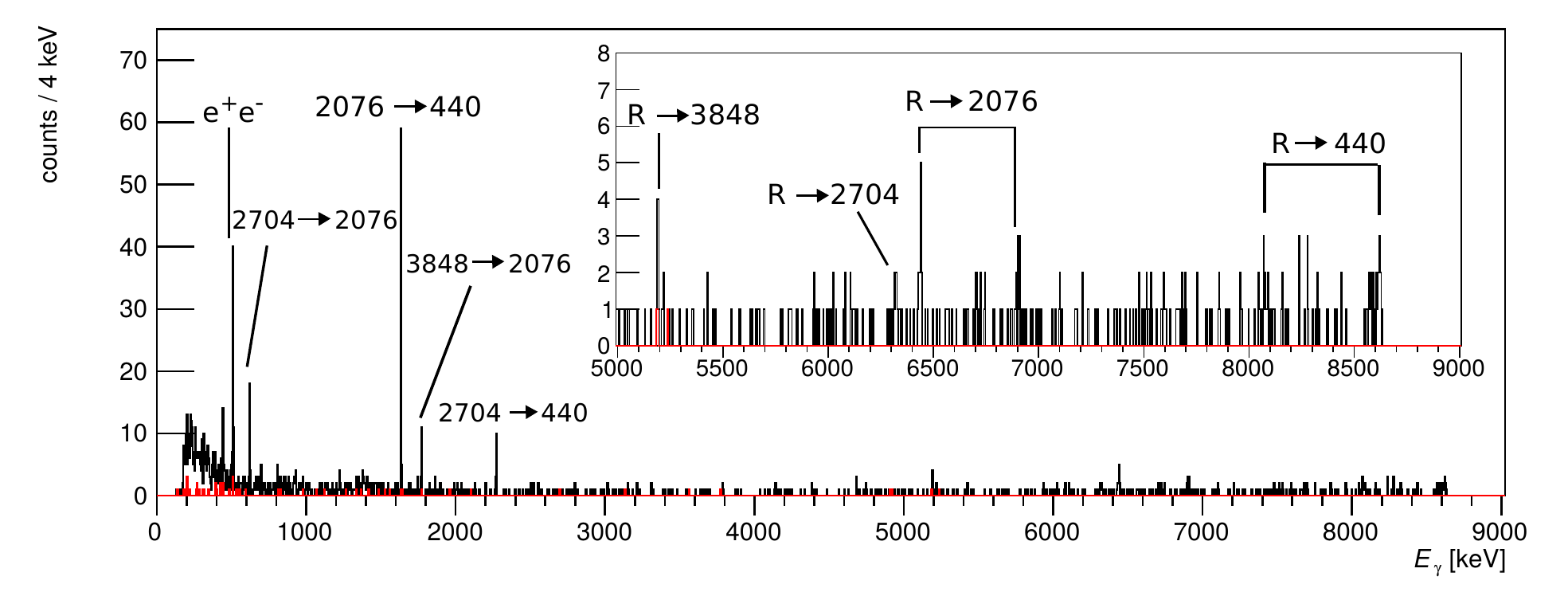}
\caption{\label{fig:259keV_spec} (Color online) Top: $\gamma$-ray spectra measured at the proton energy of maximum yield of the 259.7 keV resonance (black) and 9 keV below the maximum (turquoise). Bottom: coincidence spectrum obtained gating on the 440 keV peak in Ge90 (black) and on a background region adjacent to the 440 keV peak (red). The inset enlarges the high-energy part of the spectrum.}
\end{figure*}

The newly discovered transitions from the 8975.3 keV state to the 440 keV and 2076 keV states, which taken together account for 90$\%$ of the total decay probability, are both observed in coincidence with the 440 keV peak at 95$\%$ confidence level. The previously known \cite{Jenkins13-PRC} decay to the 2982 keV state was also observed here, but it is much weaker than the two new decays, at 3.7$\%$ branching ratio.

A resonance strength of $(1.87 \pm 0.06)\times 10^{-6}$ eV is finally obtained here for the 189.5 keV resonance. This is only slightly below the previous experimental upper limit of $\leq$2.6$\times$10$^{-6}$\,eV \cite{Goerres82-NPA}. In the previous indirect work by Hale {\it et al.} \cite{Hale01-PRC}, a value of 3.4$\times$10$^{-6}$\,eV  had been found based on the ($^3$He,d) spectroscopic factor. However, instead of their own indirect value, Hale {\it et al.} (and, subsequently, also Iliadis/STARLIB) adopted the similar experimental upper limit, $\leq$2.6$\times$10$^{-6}$\,eV \cite{Goerres82-NPA}.

\subsection{259.7 keV resonance ($E_{\rm x}$ = 9042.4 keV)}
\label{subsec:259}
The resonance at 259.7 keV is the most intense one observed in the present experiment. The resonance scan  is shown in figure \ref{fig:scan}, and a partial $\gamma$ ray spectrum with the most intense transitions is shown in figure \ref{fig:259keV_spec}.

This resonance corresponds to the 9042.4\,keV excited state in $^{23}$Na. This level was reported as part of a doublet with a 9038\,keV level \cite{Jenkins13-PRC}. According to \cite{Jenkins13-PRC}, the 9038 keV level has J$^{\pi}$\,=\,15/2$^+$, while the 9042.4\,keV level has J$^{\pi}$\,=\,7/2$^+$ or 9/2$^+$.  The $\gamma$ transitions reported in \cite{Jenkins13-PRC} for the 9038\,keV level (9038$\rightarrow$7267, 9038$\rightarrow$6234, 9038$\rightarrow$5533) are not observed here, in an almost background free spectrum (fig. \ref{fig:259keV_spec}). This is expected, because of the high angular momentum barrier due to the 15/2$^+$ angular momentum of this level. 

For the higher level in the doublet, 9042.4 (reported as 9041 in Ref. \cite{Jenkins13-PRC}, 7/2$^+$ or 9/2$^+$), the angular momentum barrier is much lower. Two decay modes were observed in the literature, 9041$\rightarrow$2076 and 9041$\rightarrow$440 \cite{Jenkins13-PRC}. These transitions correspond to the most intense $\gamma$ rays observed here. Several additional transitions have been observed in the present experiment (fig.\,\ref{fig:259keV_spec} and tab.\,\ref{tab:branching}).

Following the procedure described for the case of the 189.5 keV resonance, also here the $\gamma$-$\gamma$ coincidences were studied. The transitions from the resonance to the 440, 2076, 2703.5, 3848 and 6042 keV states are observed in coincidence with the 440 keV $\gamma$ ray (fig. \ref{fig:259keV_spec}) at 95$\%$ confidence level, confirming the correctness of the branching ratios derived in the present work.

The new resonance strength, $\omega\gamma$ = (6.89$\pm$0.16)\,$\times$\,10$^{-6}$\,eV, is twice as high as the previous experimental upper limit, $\leq$2.6$\times$\,10$^{-6}$\,eV \cite{Goerres82-NPA}. It is possible that it was missed in that work  \cite{Goerres82-NPA} due to an insufficient number of energy steps. The previous indirect upper limit is $\leq$0.13$\times$\,10$^{-6}$\,eV \cite{Hale01-PRC}, but the spin-parity was only tentatively identified there and no complete angular distribution was available, so it is possible that the measured upper limit for the spectroscopic factor referred to the other state of the doublet, the 15/2$^+$ level at 9038\,keV, which is astrophysically irrelevant due to its higher spin and, hence, higher angular momentum barrier.

\begin{table}[b!!]
\centering
\begin{ruledtabular}
\begin{tabular}{rlll}
Final level  & \multicolumn{3}{c}{Initial level [keV]} \\
\ [keV] & 8943.5 & 8975.3 & 9042.4 \\ \hline
440 &  & 42.8(0.9) & 45.4 (0.9) \\
2076 & & 47.9 (0.9) & 18.7 (0.6) \\
2391 & 23 (4) & &  \\ 
2704 & & & 10.9 (0.5) \\
2982 & & 3.7 (0.5) & \\
3848 & & & 13.3 (0.5) \\
3914 & 77 (4) & 1.1 (0.3) & 1.8 (0.4) \\
4775 & & 1.8 (0.2) & \\
5927 & & & 3.6 (0.2) \\
6042 & & & 2.6 (0.2) \\
6355 & &  & 1.5 (0.2) \\
6618 & & 2.7 (0.2) &  \\
6820 &  & & 2.2 (0.2) \\
\end{tabular}
\end{ruledtabular}
\caption{Branching ratios for the $\gamma$ decay of the $E_{\rm p}^{\rm res}$ = 156.2, 189.5, and 259.7\,keV resonances, corresponding to the $E_{\rm x}$ = 8943.5, 8975.3, and 9402.4\,keV excited states of $^{23}$Na, respectively. Statistical uncertainties are given in parentheses.}
\label{tab:branching}
\end{table}

\begin{figure}[tb]
\includegraphics[width=\columnwidth]{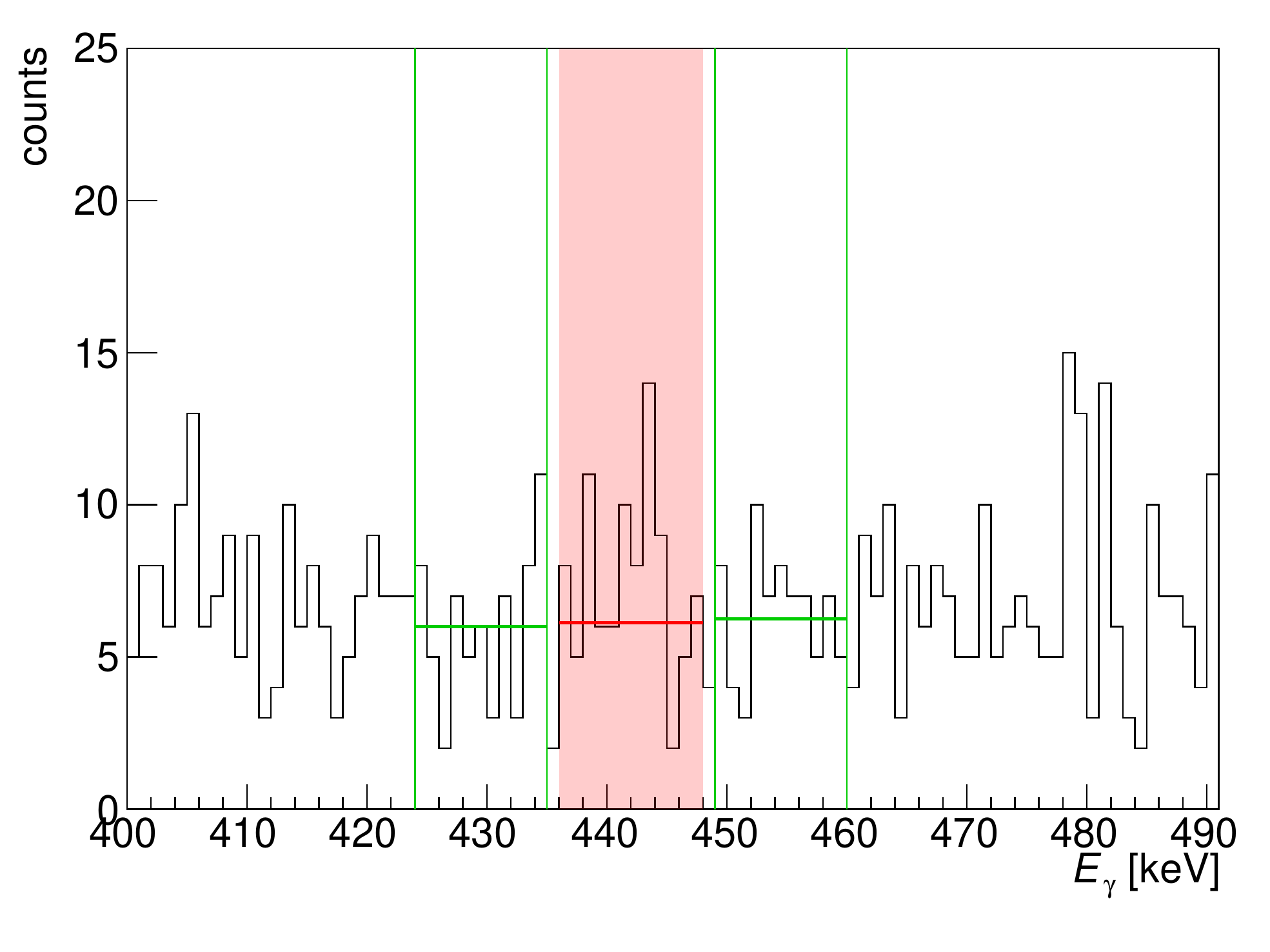}
\caption{\label{fig:ul_spec} (Color online) Regions of interest for the 440 keV peak (pink) and for the background estimation (light green lines) in the spectrum taken on the tentative 215 keV resonance.}
\end{figure}

\subsection{71 keV, 105 keV and 215 keV resonances ($E_{\rm x}$ = 8862, 8895 and 9000 keV)}
\label{subsec:ul}

The tentative resonances at 71, 105 and 215 keV were searched for by setting the beam energy to populate each resonance at the target position where both detectors had maximum efficiency. Because of the width of the efficiency plateau, for a given resonance energy $E_{p}^{\rm res}$ this search covers a beam energy range of [$E_{p}^{\rm res}$-2\,keV;$E_{p}^{\rm res}$+2\,keV].  

The existence of the three excited states corresponding to these states had been reported as tentative in an early $^{22}$Ne($^3$He,d)$^{23}$Na experiment \cite{Powers71-PRC}. A later study with the same method re-investigated the 71 and 105 keV resonances but could only find upper limits \cite{Hale01-PRC}. The Nuclear Data Sheet lists the existence of all three levels as tentative \cite{Firestone07-NDS23}. However, for each of the first two levels, corresponding to the 71 and 105 keV resonances, a definitive spin-parity assignment of $J^\Pi$ = 1/2$^+$ is given \cite{Firestone07-NDS23}. This value is justified \cite{Firestone07-NDS23} with the assumption of $l$=0 angular momentum transfer in the $^{22}$Ne($^3$He,d)$^{23}$Na reaction. However, these levels were only tentatively or not at all detected in the two $^{22}$Ne($^3$He,d)$^{23}$Na experiments \cite{Powers71-PRC,Hale01-PRC}, so the assumed $J^\Pi$ assignment for the 71 and 105 keV resonances \cite{Firestone07-NDS23} seems speculative and is not followed here. For the tentative 215 keV resonance, no spin-parity assignment is given \cite{Firestone07-NDS23}.

Also in the present experiment, none of the three resonances was detected. Instead, upper limits for the resonance strengths were derived assuming $100\%$ branching ratio for the 440$\rightarrow$0 keV transition from the first excited to the ground state.  

For the evaluation of the upper limits the Rolke method has been used \cite{Rolke05-NIM}, which applies the profile likelihood technique in the case of a Poisson-distributed signal, taking into account the uncertainty on the background determination. In the present case the background estimate is based on experimental data, i.e. on the number of counts in two spectral regions on the left and on the right side of the 440 keV peak. The background thus has similar uncertainty as the signal, which is why the Rolke method is used here.
Fig. \ref{fig:ul_spec} shows the regions of interest adopted for the signal and for the background estimation in the case of the tentative 215 keV resonance. The regions of interest were chosen using  spectra from different, observed resonances where the 440 keV peak was clearly in evidence, and then used without change for the tentative resonances.

\begin{table*}[t!!]
\caption{$^{22}$Ne(p,$\gamma$)$^{23}$Na resonance strengths used for the reaction rate calculation. Upper limits from the present work are given at 90\% confidence level. Resonances at $E_p >$ 661\,keV are adopted without change from Ref.~\cite{Sallaska13-ApJSS} but not listed here. See text for details.}
\begin{center}
\label{tab:wg_TRR}
\begin{tabular}{ d | c | r | r | r | c }
\hline
\multicolumn{1}{c|}{\textbf{$E_{\rm p}^{\rm res}$}} & \multicolumn{4}{c|}{\textbf{Strength} $\omega\gamma$\,[eV]} & \multicolumn{1}{c}{Screening} \\
\multicolumn{1}{c|}{$[$keV$]$} & \multicolumn{1}{c|}{Literature} & \multicolumn{1}{c|}{Literature} & \multicolumn{1}{c|}{Present work,} & \multicolumn{1}{c|}{Adopted} & \multicolumn{1}{c}{enhancement} \\
   & direct & \multicolumn{1}{c|}{indirect} & \multicolumn{1}{c|}{and Ref. \cite{Cavanna15-PRL}} & & \multicolumn{1}{c}{factor $f$} \\
\hline
\hline
29 & - & $\leq$\,2.6\,$\times$\,10$^{-25}$\,\cite{Iliadis10-NPA841_251} & - & $\leq$\,2.6\,$\times$\,10$^{-25}$ \\
37 & - & (3.1\,$\pm$\,1.2)\,$\times$\,10$^{-15}$\,\cite{Iliadis10-NPA841_251} & - & (3.1\,$\pm$\,1.2)\,$\times$\,10$^{-15}$ \\ 
71 & $\leq$\,3.2\,$\times$\,10$^{-6}$\,\cite{Goerres82-NPA} & $\leq$\,1.9\,$\times$\,10$^{-10}$\,\cite{Hale01-PRC} & $\leq$\,1.5\,$\times$\,10$^{-9}$ & $\leq$\,1.5\,$\times$\,10$^{-9}$ & 1.266\\
105 & $\leq$\,0.6\,$\times$\,10$^{-6}$\,\cite{Goerres82-NPA} & $\leq$\,1.4\,$\times$\,10$^{-7}$\,\cite{Hale01-PRC} & $\leq$\,7.6\,$\times$\,10$^{-9}$ & $\leq$\,7.6\,$\times$\,10$^{-9}$ & 1.140\\
156.2 & $\leq$\,1.0\,$\times$\,10$^{-6}$\,\cite{Goerres82-NPA} & (9.2\,$\pm$\,3.7)\,$\times$\,10$^{-9}$\,\cite{Iliadis10-NPA841_251} & (1.48\,$\pm$\,0.09$_{stat}$\,$\pm$\,0.04$_{\rm syst}$)\,$\times$\,10$^{-7}$ & (1.48\,$\pm$\,0.10)\,$\times$\,10$^{-7}$ & 1.074\\
189.5 & $\leq$\,2.6\,$\times$\,10$^{-6}$\,\cite{Goerres82-NPA} & 3.4\,$\times$\,10$^{-6}$\,\cite{Hale01-PRC} & (1.87\,$\pm$\,0.03$_{stat}$\,$\pm$\,0.05$_{\rm syst}$)\,$\times$\,10$^{-6}$ & (1.87\,$\pm$\,0.06)\,$\times$\,10$^{-6}$ & 1.055\\
215 & $\leq$\,1.4\,$\times$\,10$^{-6}$\,\cite{Goerres82-NPA} & - & $\leq$\,2.8\,$\times$\,10$^{-8}$ & $\leq$\,2.8\,$\times$\,10$^{-8}$ & 1.045\\
259.7 & $\leq$\,2.6\,$\times$\,10$^{-6}$\,\cite{Goerres82-NPA} & $\leq$\,1.3\,$\times$\,10$^{-7}$\,\cite{Hale01-PRC} & (6.89\,$\pm$\,0.07$_{stat}$\,$\pm$\,0.15$_{\rm syst}$)\,$\times$\,10$^{-6}$ & (6.89\,$\pm$\,0.16)\,$\times$\,10$^{-6}$ & 1.034\\
291 & $\leq$\,2.2\,$\times$\,10$^{-6}$\,\cite{Goerres82-NPA} & - & - & $\leq$\,2.2\,$\times$\,10$^{-6}$ \\
323 & $\leq$\,2.2\,$\times$\,10$^{-6}$\,\cite{Goerres82-NPA} & - & - & $\leq$\,2.2\,$\times$\,10$^{-6}$ \\
334 & $\leq$\,3.0\,$\times$\,10$^{-6}$\,\cite{Goerres82-NPA} & - & - & $\leq$\,3.0\,$\times$\,10$^{-6}$ \\
369 & - & $\leq$\,6.0\,$\times$\,10$^{-4}$\,\cite{Hale01-PRC} & - & $\leq$\,6.0\,$\times$\,10$^{-4}$ \\
394 & - & $\leq$\,6.0\,$\times$\,10$^{-4}$\,\cite{Hale01-PRC} & - & $\leq$\,6.0\,$\times$\,10$^{-4}$\\
436 & 0.079 $\pm$ 0.006 \cite{Depalo15-PRC}  & & - & 0.079 $\pm$ 0.006\\
479 & 0.594 $\pm$ 0.038 \cite{Depalo15-PRC} & & - & 0.594 $\pm$ 0.038\\
638.5 & 2.45 $\pm$ 0.18 \cite{Depalo15-PRC} & - & & 2.45 $\pm$ 0.18\\
661 & 0.032 $\pm$ 0.015 \cite{Depalo15-PRC} & & - & 0.032 $\pm$ 0.015\\
\hline          
\end{tabular}
\end{center}
\end{table*}%

The present new upper limits are summarized in table \ref{tab:wg_TRR}. Owing to the ultra-low background at LUNA, the new limits are orders of magnitude lower than the upper limits from the previous direct experiment \cite{Goerres82-NPA}. 

It should be noted that the limits calculated are purely experimental, different from indirect upper limits available from the literature which depend on the spin-parity assumed and on the normalization for spectroscopic factors \cite{Hale01-PRC}.

\section{Astrophysical aspects}
\label{sec:Astro}

\begin{figure*}[t!!]
\includegraphics[width=\textwidth]{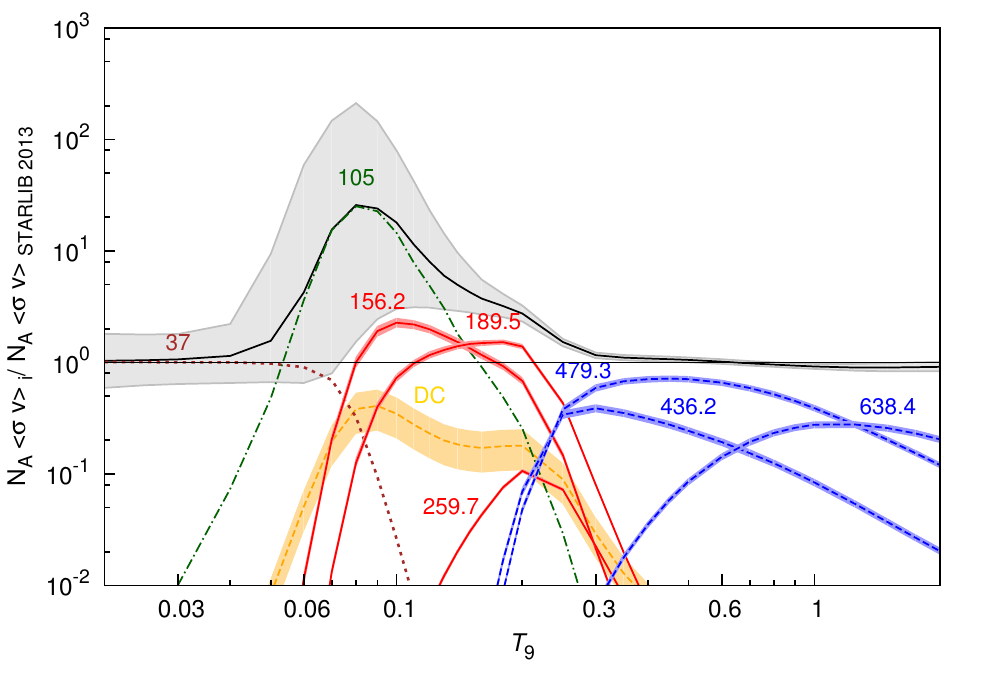}
\caption{\label{fig:rate_allreso} (Color online) $^{22}$Ne(p,$\gamma$)$^{23}$Na total reaction rate normalized to STARLIB 2013 \cite{Sallaska13-ApJSS} (black full line), and its 1$\sigma$ error bar (grey band). The contribution of individual resonances and of the direct capture component is given in color, restricted to resonances contributing more than 1$\%$ to the previous total rate. }
\end{figure*}

An updated reaction rate has been calculated using the results from the present experiment and the most recent literature data \cite{Depalo15-PRC,Sallaska13-ApJSS}. The input parameters for the reaction rate calculation are listed in table \ref{tab:wg_TRR} and explained in the following. 

The effect of cross section enhancement due to electron screening \cite{Assenbaum87-ZPA} has been taken into account in the reaction rate calculation. 
The resonance strength is proportional to the proton partial width $\Gamma_p$, which, in turn, depends on the Coulomb barrier penetrability $P_l(E)$ as a function of partial wave $l$ and energy $E$. The enhancement of the cross section, hence the resonance strength, measured in the laboratory due to the electron screening effect  \cite{Assenbaum87-ZPA} can be expressed by the enhancement factor $f$ given by:
\begin{equation}
f = \frac{\omega\gamma_{\rm measured}}{\omega\gamma_{\rm bare}} = \frac{P_l(E+U_e)}{P_l(E)}
\label{eq:screening_1}
\end{equation} 
where $E$ is the resonance energy in the center-of-mass system, $U_e$ is the screening potential, and $\omega\gamma_{\rm bare}$ is the resonance strength for a bare nucleus without atomic electrons. In the adiabatic limit, the screening potential $U_e$ is given by the difference between the total electron binding energy of neutral $^{23}$Na and the combined binding energies of neutral $^{22}$Ne and neutral $^{1}$H \cite{Assenbaum87-ZPA}. Using tabulated values \cite{Huang76-ADNDT}, $U_e$ = 0.895 keV is found. For $l$ = 0, which is usually the case at low energy, $f$ can be approximated as \cite{Clayton84-Book}:
\begin{equation}
f = \left( \frac{E}{E+U_e} \right)^{1/2} \exp\left[-2\pi\eta(E+U_e) + 2\pi\eta(E)\right]
\label{eq:screening_2}
\end{equation} 
where $\eta(E)$ is the Sommerfeld parameter \cite{Iliadis07-Book}.
For the resonances studied here, the calculated screening enhancement factors $f$ in the adiabatic limit are included in table \ref{tab:wg_TRR}. In order to obtain the bare resonance strength $\omega\gamma_{\rm bare}$, the strength values listed in table \ref{tab:wg_TRR} have to be divided by the respective $f$ value.

For the intermediate-energy resonances at $E_p$ = 436, 479, 638.5, and 661\,keV, recent experimental data  \cite{Depalo15-PRC} have been used. For the resonances at even higher energies, $E_p > $ 661 keV (not listed in table \ref{tab:wg_TRR}), and for the 29 and 37 keV resonances, energy and strength data from the STARLIB compilation \cite{Sallaska13-ApJSS} are used here. 

Using these input data, a Monte Carlo technique has been adopted to calculate the reaction rate: a random sampling of the input parameters that enter the reaction rate calculation was performed, assuming a Gaussian distribution for the resonance energy, and a lognormal distribution for the resonance strength \cite{Longland10-NPA841_1}. 

The upper limits on the strengths of the tentative resonances at 71, 105 and 215 keV were also included using a Monte Carlo approach but taking into account the probability density function of the present experimental data: for each tentative resonance, Poisson-distributed random counts in the region of interest and in the background were extracted, then a number of net counts was derived subtracting the background from the ROI counts. The resonance strength was then calculated using eq. (\ref{eq:wg}). 

For each of the temperatures in table \ref{tab:TRR}, the random sampling of all parameters was repeated 10$^4$ times and the total resonant reaction rate $N_{\rm A}\langle \sigma v\rangle_{\rm R}$ was calculated for a grid of temperatures $T_9$ with the relationship \cite{Iliadis07-Book}:  
\begin{equation}
\label{eq:tnrr}
N_{\rm A}\langle \sigma v\rangle_{\rm R}\,=\,\frac{1.5399\,10^{5}}{(\mu T_9)^{3/2}} \displaystyle\sum_i (\omega\gamma)_i \exp[-11.605 E^{\rm res}_{i}/T_9]
\end{equation}
where $T_9$ is the temperature in GK, $\mu=m_{22}m_p/(m_{22}+m_p)$ the reduced mass in amu (atomic mass units), $(\omega\gamma)_i$ the strength of resonance $i$ in eV, and $E^{\rm res}_{i}$ the center-of-mass energy of resonance $i$ in MeV. The total reaction rate $N_{\rm A}\langle \sigma v\rangle$ is then given by:
\begin{equation}
\label{eq:tnrrtot}
N_{\rm A}\langle \sigma v\rangle\,=\,N_{\rm A}\langle \sigma v\rangle_{\rm R} + N_{\rm A}\langle \sigma v\rangle_{\rm DC}
\end{equation}
with $N_{\rm A}\langle \sigma v\rangle_{\rm DC}$ the direct capture contribution, assuming a constant $S$-factor of 62 keV$\cdot$b with a relative uncertainty of 40\% \cite{Goerres83-NPA}. 

At each given temperature, the `low', `median' and `high' value of the thermonuclear reaction rate were then determined using the 0.16, 0.50, and 0.84 quantiles of the probability density function for the total reaction rate. 

The three tentative resonances at 71, 105 and 215 keV were treated in a special manner for the `low' rate. For this rate, they were simply set to zero, with uncertainty zero, as had been done previously by Iliadis/STARLIB \cite{Iliadis10-NPA841_31,Iliadis10-NPA841_251,Sallaska13-ApJSS}. In this way, the remaining uncertainty due to the non-zero experimental upper limit led to a relatively large error band especially at $T_9$ $\sim$ 0.07, where the influence of the 105 keV resonance is greatest. This error is much larger than the one by Iliadis/STARLIB \cite{Iliadis10-NPA841_31, Sallaska13-ApJSS}, but more realistic given the strong differences between indirect and direct results observed here for the 156.2 and 259.7 keV resonances.

The reaction rate is provided in tabular form in table \ref{tab:TRR}. In the temperature range 0.02$\leq$$T_9$$\leq$2, it can be approximated to $\pm$4\% by the following analytical formula:
\begin{eqnarray}
\label{rate_approx}
N_{\rm A}\langle \sigma v\rangle & = & p_0 T_9^{-2/3} \exp (p_1 T_9^{-1/3}) \\
 & &  + p_2 T_9^{-3/2} \exp (p_3/T_9) + p_4 T_9^{-3/2} \exp (p_5/T_9) \nonumber \\
 & &  + p_6 T_9^{-3/2} \exp (p_7/T_9) + p_8 T_9^{-1/2} \exp (p_{9}/T_9) \nonumber \\ 
 & & + p_{10} T_9^{-3/2} \exp (p_{11}/T_9) \nonumber
\end{eqnarray}
with coefficients:
\begin{eqnarray}
\label{rate_approx_coeff}
p_{0}= 1.086\times10^{+6} & \qquad & 
p_{1}= -1.550\times10^{+1}\nonumber \\
p_{2}= 5.073\times10^{-10} & & 
p_{3}= -4.120\times10^{-1}\nonumber \\
p_{4}= 4.394\times10^{-1} & & 
p_{5}= -2.097 \nonumber \\
p_{6}= 1.031\times10^{+5} & & 
p_{7}= -5.178 \nonumber \\
p_{8}= 8.327\times10^{+5} & & 
p_{9}= -7.343 \nonumber \\
p_{10}= 9.198\times10^{-4} & & 
p_{11}= -1.229 \nonumber \\
\end{eqnarray}

The contribution of the various resonances to the total reaction rate is shown in fig. \ref{fig:rate_allreso} and discussed in the following paragraphs.

At the very lowest temperatures, $T_9$ $\leq$ 0.05, the reaction rate is almost completely dominated by the 37\,keV resonance, whose energy and strength are so low that a direct strength measurement cannot be conceived of with present technology. It has to be noted that at these low temperatures, the NeNa cycle is very inefficient and hardly contributes to nucleosynthesis, when compared to the CNO cycles.

At somewhat higher temperatures, $T_9$ = 0.05-0.1, the present median rate is dominated by the tentative resonance at 105\,keV. Even though it is not observed, the little rate contributed by the nonzero part of the probability distribution function derived from the present upper limit is enough to dominate the recommended rate, and the error bar. It is clear that this resonance warrants further study. Just such a measurement is presently underway at LUNA. It should be noted that the 1$\sigma$ lower limit of the thermonuclear reaction rate has been calculated setting this resonance to zero.

At even higher temperatures, $T_9$ = 0.08-0.2, each of the two newly observed resonances at 156.2 and 189.5 keV contributes more to the rate than the total recommended rate in the previous Iliadis/STARLIB compilations. The total rate enhancement is up to a factor of three, even when the tentative resonance at 105 keV is left out. The relatively strong new resonance at 259.7 keV contributes much less to the reaction rate, only up to 10\% of the previous total. The direct capture component, assumed to be described by a constant $S$-factor of 62\,keV\,barn \cite{Goerres83-NPA}, contributes in this temperature range, as well.

For even higher temperatures, 0.2 $\leq$ $T_9$, a number of resonances above 400 keV play a role, and in particular three resonances recently studied at HZDR \cite{Depalo15-PRC} are included in the present rate.

The tentative resonances at 71 and 215 keV have been omitted from fig. \ref{fig:rate_allreso}, because they contribute less than 1\% to the previous total rate at all temperatures shown here.

\begin{table}[tbh]
\centering
\begin{ruledtabular}
\begin{tabular}{ d l l l}
\multicolumn{1}{c}{$T_9$} & Low Rate & Median Rate & High Rate \\
\hline
0.01  &   2.936$\times10^{-25 }$ &  6.915$\times10^{-25 }$ &  1.643$\times10^{-24 }$ \\
0.011  &  1.135$\times10^{-23 }$ &  2.553$\times10^{-23 }$ &  5.756$\times10^{-23 }$ \\
0.012  &  2.402$\times10^{-22 }$ &  5.058$\times10^{-22 }$ &  1.059$\times10^{-21 }$ \\
0.013  &  3.036$\times10^{-21 }$ &  6.088$\times10^{-21 }$ &  1.226$\times10^{-20 }$ \\
0.014  &  2.671$\times10^{-20 }$ &  5.233$\times10^{-20 }$ &  1.023$\times10^{-19 }$ \\
0.015  &  1.762$\times10^{-19 }$ &  3.338$\times10^{-19 }$ &  6.471$\times10^{-19 }$ \\
0.016  &  9.133$\times10^{-19 }$ &  1.687$\times10^{-18 }$ &  3.092$\times10^{-18 }$ \\
0.018  &  1.358$\times10^{-17 }$ &  2.463$\times10^{-17 }$ &  4.432$\times10^{-17 }$ \\
0.02  &   1.173$\times10^{-16 }$ &  2.053$\times10^{-16 }$ &  3.576$\times10^{-16 }$ \\
0.025  &  5.423$\times10^{-15 }$ &  9.089$\times10^{-15 }$ &  1.545$\times10^{-14 }$ \\
0.03  &   6.538$\times10^{-14 }$ &  1.086$\times10^{-13 }$ &  1.836$\times10^{-13 }$ \\
0.04  &   1.335$\times10^{-12 }$ &  2.337$\times10^{-12 }$ &  4.519$\times10^{-12 }$ \\
0.05  &   7.723$\times10^{-12 }$ &  1.811$\times10^{-11 }$ &  1.094$\times10^{-10 }$ \\
0.06  &   2.450$\times10^{-11 }$ &  1.589$\times10^{-10 }$ &  2.209$\times10^{-9 }$ \\
0.07  &   8.370$\times10^{-11 }$ &  1.631$\times10^{-9 }$ &  1.550$\times10^{-8 }$ \\
0.08  &   5.892$\times10^{-10 }$ &  9.926$\times10^{-9 }$ &  8.124$\times10^{-8 }$ \\
0.09  &   4.624$\times10^{-9 }$ &  4.544$\times10^{-8 }$ &  2.744$\times10^{-7 }$ \\
0.1  &    2.783$\times10^{-8 }$ &  1.654$\times10^{-7 }$ &  7.338$\times10^{-7 }$ \\
0.11  &   1.257$\times10^{-7 }$ &  4.575$\times10^{-7 }$ &  1.695$\times10^{-6 }$ \\
0.12  &   4.503$\times10^{-7 }$ &  1.160$\times10^{-6 }$ &  3.348$\times10^{-6 }$ \\
0.13  &   1.337$\times10^{-6 }$ &  2.690$\times10^{-6 }$ &  6.410$\times10^{-6 }$ \\
0.14  &   3.413$\times10^{-6 }$ &  5.857$\times10^{-6 }$ &  1.135$\times10^{-5 }$ \\
0.15  &   7.740$\times10^{-6 }$ &  1.173$\times10^{-5 }$ &  1.980$\times10^{-5 }$ \\
0.16  &   1.585$\times10^{-5 }$ &  2.205$\times10^{-5 }$ &  3.259$\times10^{-5 }$ \\
0.18  &   5.327$\times10^{-5 }$ &  6.688$\times10^{-5 }$ &  8.602$\times10^{-5 }$ \\
0.2  &    1.466$\times10^{-4 }$ &  1.732$\times10^{-4 }$ &  2.038$\times10^{-4 }$ \\
0.25  &   1.653$\times10^{-3 }$ &  1.797$\times10^{-3 }$ &  1.917$\times10^{-3 }$ \\
0.3  &    2.160$\times10^{-2 }$ &  2.295$\times10^{-2 }$ &  2.429$\times10^{-2 }$ \\
0.35  &   1.807$\times10^{-1 }$ &  1.910$\times10^{-1 }$ &  2.015$\times10^{-1 }$ \\
0.4  &    9.285$\times10^{-1 }$ &  9.793$\times10^{-1 }$ &  1.032$\times10^{+0 }$ \\
0.45  &   3.332$\times10^{+0 }$ &  3.510$\times10^{+0 }$ &  3.697$\times10^{+0 }$ \\
0.5  &    9.281$\times10^{+0 }$ &  9.767$\times10^{+0 }$ &  1.028$\times10^{+1 }$ \\
0.6  &    4.335$\times10^{+1 }$ &  4.549$\times10^{+1 }$ &  4.773$\times10^{+1 }$ \\
0.7  &    1.320$\times10^{+2 }$ &  1.382$\times10^{+2 }$ &  1.448$\times10^{+2 }$ \\
0.8  &    3.088$\times10^{+2 }$ &  3.232$\times10^{+2 }$ &  3.383$\times10^{+2 }$ \\
0.9  &    6.067$\times10^{+2 }$ &  6.366$\times10^{+2 }$ &  6.680$\times10^{+2 }$ \\
1  &  	  1.059$\times10^{+3 }$ &  1.112$\times10^{+3 }$ &  1.175$\times10^{+3 }$ \\
1.25  &   2.992$\times10^{+3 }$ &  3.189$\times10^{+3 }$ &  3.430$\times10^{+3 }$ \\
1.5  &    6.240$\times10^{+3 }$ &  6.749$\times10^{+3 }$ &  7.373$\times10^{+3 }$ \\
1.75  &   1.081$\times10^{+4 }$ &  1.179$\times10^{+4 }$ &  1.301$\times10^{+4 }$ \\
2  &      1.657$\times10^{+4 }$ &  1.814$\times10^{+4 }$ &  2.014$\times10^{+4 }$ \\
2.5  &    3.046$\times10^{+4 }$ &  3.342$\times10^{+4 }$ &  3.729$\times10^{+4 }$ \\
3  &  	  4.585$\times10^{+4 }$ &  5.032$\times10^{+4 }$ &  5.586$\times10^{+4 }$ \\
3.5  &    6.113$\times10^{+4 }$ &  6.689$\times10^{+4 }$ &  7.397$\times10^{+4 }$ \\
4  &  	  7.526$\times10^{+4 }$ &  8.187$\times10^{+4 }$ &  8.967$\times10^{+4 }$ \\
\end{tabular}
\end{ruledtabular}
\caption{Thermonuclear reaction rate $N_{\rm A}\langle \sigma v\rangle$ for $^{22}$Ne(p,$\gamma$)$^{23}$Na in units cm$^{-3}$s$^{-1}$mol$^{-1}$, as a function of temperature $T_9$.}
\label{tab:TRR}
\end{table}

\section{Summary and outlook}
\label{sec:Summary}
A new direct study of the $^{22}$Ne(p,$\gamma$)$^{23}$Na reaction has been performed deep underground at LUNA. Three resonances at 156.2 keV, 189.5 keV and 259.7 keV have been observed for the first time. For these resonances, new resonance strengths $\omega\gamma$ have been measured, superseding the previous upper limits. Moreover, new $\gamma$-ray transitions and the corresponding branching ratios are provided. Two of the three new resonances observed here (156.2 and 259.7 keV) have an experimental strength that is more than a factor of ten higher than a previous indirect upper limit, underlining the uncertainties involved when using indirect data.

The three new resonances lie directly in the Gamow window for the hot bottom burning process in AGB stars, and they also address the lower part of the Gamow window for classical novae explosions. The full astrophysical implications will be addressed in a forthcoming publication.

Improved, direct upper limits for three tentative resonances at 71, 105 and 215 keV have also been determined. The new limits are two to three orders of magnitude lower than the limits from previous work and do not rely on spin-parity assumptions that may significantly affect the upper limits determined with indirect approaches.

A new study of the $^{22}$Ne(p,$\gamma$)$^{23}$Na reaction is currently ongoing at LUNA using a high-efficiency $4\pi$ BGO detector. In that experiment, the tentative resonances and the direct capture contribution will be investigated with unprecedented sensitivity. 

\begin{acknowledgments}
Financial support by INFN, DFG (BE 4100-2/1), NAVI (HGF VH-VI-417), OTKA (K120666), as well as DAAD fellowships at HZDR for R.D. and F.C. are gratefully acknowledged. 
\end{acknowledgments}


\begin{thebibliography}{44}%
\makeatletter
\providecommand \@ifxundefined [1]{%
 \@ifx{#1\undefined}
}%
\providecommand \@ifnum [1]{%
 \ifnum #1\expandafter \@firstoftwo
 \else \expandafter \@secondoftwo
 \fi
}%
\providecommand \@ifx [1]{%
 \ifx #1\expandafter \@firstoftwo
 \else \expandafter \@secondoftwo
 \fi
}%
\providecommand \natexlab [1]{#1}%
\providecommand \enquote  [1]{``#1''}%
\providecommand \bibnamefont  [1]{#1}%
\providecommand \bibfnamefont [1]{#1}%
\providecommand \citenamefont [1]{#1}%
\providecommand \href@noop [0]{\@secondoftwo}%
\providecommand \href [0]{\begingroup \@sanitize@url \@href}%
\providecommand \@href[1]{\@@startlink{#1}\@@href}%
\providecommand \@@href[1]{\endgroup#1\@@endlink}%
\providecommand \@sanitize@url [0]{\catcode `\\12\catcode `\$12\catcode
  `\&12\catcode `\#12\catcode `\^12\catcode `\_12\catcode `\%12\relax}%
\providecommand \@@startlink[1]{}%
\providecommand \@@endlink[0]{}%
\providecommand \url  [0]{\begingroup\@sanitize@url \@url }%
\providecommand \@url [1]{\endgroup\@href {#1}{\urlprefix }}%
\providecommand \urlprefix  [0]{URL }%
\providecommand \Eprint [0]{\href }%
\providecommand \doibase [0]{http://dx.doi.org/}%
\providecommand \selectlanguage [0]{\@gobble}%
\providecommand \bibinfo  [0]{\@secondoftwo}%
\providecommand \bibfield  [0]{\@secondoftwo}%
\providecommand \translation [1]{[#1]}%
\providecommand \BibitemOpen [0]{}%
\providecommand \bibitemStop [0]{}%
\providecommand \bibitemNoStop [0]{.\EOS\space}%
\providecommand \EOS [0]{\spacefactor3000\relax}%
\providecommand \BibitemShut  [1]{\csname bibitem#1\endcsname}%
\let\auto@bib@innerbib\@empty
\bibitem [{\citenamefont {Marion}\ and\ \citenamefont
  {Fowler}(1957)}]{Marion57-ApJ}%
  \BibitemOpen
  \bibfield  {author} {\bibinfo {author} {\bibfnamefont {J.}~\bibnamefont
  {Marion}}\ and\ \bibinfo {author} {\bibfnamefont {W.}~\bibnamefont
  {Fowler}},\ }\href {\doibase 10.1086/146296} {\bibfield  {journal} {\bibinfo
  {journal} {Astrophys.~J.}\ }\textbf {\bibinfo {volume} {125}},\ \bibinfo
  {pages} {221} (\bibinfo {year} {1957})}\BibitemShut {NoStop}%
\bibitem [{\citenamefont {{Gratton}}\ \emph {et~al.}(2012)\citenamefont
  {{Gratton}}, \citenamefont {{Carretta}},\ and\ \citenamefont
  {{Bragaglia}}}]{Gratton12-AAR}%
  \BibitemOpen
  \bibfield  {author} {\bibinfo {author} {\bibfnamefont {R.~G.}\ \bibnamefont
  {{Gratton}}}, \bibinfo {author} {\bibfnamefont {E.}~\bibnamefont
  {{Carretta}}}, \ and\ \bibinfo {author} {\bibfnamefont {A.}~\bibnamefont
  {{Bragaglia}}},\ }\href {\doibase 10.1007/s00159-012-0050-3} {\bibfield
  {journal} {\bibinfo  {journal} {Astron.~Astrophys.~Revs.}\ }\textbf {\bibinfo
  {volume} {20}},\ \bibinfo {eid} {50} (\bibinfo {year} {2012})}\BibitemShut
  {NoStop}%
\bibitem [{\citenamefont {{N. Prantzos}}\ \emph {et~al.}(2007)\citenamefont
  {{N. Prantzos}}, \citenamefont {{C. Charbonnel}},\ and\ \citenamefont {{C.
  Iliadis}}}]{Prantzos07-AA}%
  \BibitemOpen
  \bibfield  {author} {\bibinfo {author} {\bibnamefont {{N. Prantzos}}},
  \bibinfo {author} {\bibnamefont {{C. Charbonnel}}}, \ and\ \bibinfo {author}
  {\bibnamefont {{C. Iliadis}}},\ }\href {\doibase 10.1051/0004-6361:20077205}
  {\bibfield  {journal} {\bibinfo  {journal} {Astron.~Astrophys.}\ }\textbf
  {\bibinfo {volume} {470}},\ \bibinfo {pages} {179} (\bibinfo {year}
  {2007})}\BibitemShut {NoStop}%
\bibitem [{\citenamefont {Ventura}\ \emph {et~al.}(2013)\citenamefont
  {Ventura}, \citenamefont {Di~Criscienzo}, \citenamefont {Carini},\ and\
  \citenamefont {D'Antona}}]{Ventura13-MNRAS}%
  \BibitemOpen
  \bibfield  {author} {\bibinfo {author} {\bibfnamefont {P.}~\bibnamefont
  {Ventura}}, \bibinfo {author} {\bibfnamefont {M.}~\bibnamefont
  {Di~Criscienzo}}, \bibinfo {author} {\bibfnamefont {R.}~\bibnamefont
  {Carini}}, \ and\ \bibinfo {author} {\bibfnamefont {F.}~\bibnamefont
  {D'Antona}},\ }\href {\doibase 10.1093/mnras/stt444} {\bibfield  {journal}
  {\bibinfo  {journal} {Monthly Notices of the Royal Astronomical Society}\
  }\textbf {\bibinfo {volume} {431}},\ \bibinfo {pages} {3642} (\bibinfo {year}
  {2013})}\BibitemShut {NoStop}%
\bibitem [{\citenamefont {Ventura}\ \emph {et~al.}(2012)\citenamefont
  {Ventura}, \citenamefont {D'Antona}, \citenamefont {Criscienzo},
  \citenamefont {Carini}, \citenamefont {D'Ercole},\ and\ \citenamefont
  {vesperini}}]{Ventura12-ApJ}%
  \BibitemOpen
  \bibfield  {author} {\bibinfo {author} {\bibfnamefont {P.}~\bibnamefont
  {Ventura}}, \bibinfo {author} {\bibfnamefont {F.}~\bibnamefont {D'Antona}},
  \bibinfo {author} {\bibfnamefont {M.~D.}\ \bibnamefont {Criscienzo}},
  \bibinfo {author} {\bibfnamefont {R.}~\bibnamefont {Carini}}, \bibinfo
  {author} {\bibfnamefont {A.}~\bibnamefont {D'Ercole}}, \ and\ \bibinfo
  {author} {\bibfnamefont {E.}~\bibnamefont {vesperini}},\ }\href
  {http://stacks.iop.org/2041-8205/761/i=2/a=L30} {\bibfield  {journal}
  {\bibinfo  {journal} {The Astrophysical Journal Letters}\ }\textbf {\bibinfo
  {volume} {761}},\ \bibinfo {pages} {L30} (\bibinfo {year}
  {2012})}\BibitemShut {NoStop}%
\bibitem [{\citenamefont {{Doherty}}\ \emph {et~al.}(2014)\citenamefont
  {{Doherty}}, \citenamefont {{Gil-Pons}}, \citenamefont {{Lau}}, \citenamefont
  {{Lattanzio}},\ and\ \citenamefont {{Siess}}}]{Doherty14-MNRAS}%
  \BibitemOpen
  \bibfield  {author} {\bibinfo {author} {\bibfnamefont {C.~L.}\ \bibnamefont
  {{Doherty}}}, \bibinfo {author} {\bibfnamefont {P.}~\bibnamefont
  {{Gil-Pons}}}, \bibinfo {author} {\bibfnamefont {H.~H.~B.}\ \bibnamefont
  {{Lau}}}, \bibinfo {author} {\bibfnamefont {J.~C.}\ \bibnamefont
  {{Lattanzio}}}, \ and\ \bibinfo {author} {\bibfnamefont {L.}~\bibnamefont
  {{Siess}}},\ }\href {\doibase 10.1093/mnras/stt1877} {\bibfield  {journal}
  {\bibinfo  {journal} {Monthly Notices Royal Astronom. Soc.}\ }\textbf
  {\bibinfo {volume} {437}},\ \bibinfo {pages} {195} (\bibinfo {year}
  {2014})}\BibitemShut {NoStop}%
\bibitem [{\citenamefont {{Decressin}}\ \emph {et~al.}(2007)\citenamefont
  {{Decressin}}, \citenamefont {{Meynet}}, \citenamefont {{Charbonnel}},
  \citenamefont {{Prantzos}},\ and\ \citenamefont
  {{Ekstr{\"o}m}}}]{Decressin07-AA}%
  \BibitemOpen
  \bibfield  {author} {\bibinfo {author} {\bibfnamefont {T.}~\bibnamefont
  {{Decressin}}}, \bibinfo {author} {\bibfnamefont {G.}~\bibnamefont
  {{Meynet}}}, \bibinfo {author} {\bibfnamefont {C.}~\bibnamefont
  {{Charbonnel}}}, \bibinfo {author} {\bibfnamefont {N.}~\bibnamefont
  {{Prantzos}}}, \ and\ \bibinfo {author} {\bibfnamefont {S.}~\bibnamefont
  {{Ekstr{\"o}m}}},\ }\href {\doibase 10.1051/0004-6361:20066013} {\bibfield
  {journal} {\bibinfo  {journal} {Astron.~Astrophys.}\ }\textbf {\bibinfo
  {volume} {464}},\ \bibinfo {pages} {1029} (\bibinfo {year}
  {2007})}\BibitemShut {NoStop}%
\bibitem [{\citenamefont {Denissenkov}\ \emph {et~al.}(2015)\citenamefont
  {Denissenkov}, \citenamefont {VandenBerg}, \citenamefont {Hartwick},
  \citenamefont {Herwig}, \citenamefont {Weiss},\ and\ \citenamefont
  {Paxton}}]{Denissenkov15-MNRAS}%
  \BibitemOpen
  \bibfield  {author} {\bibinfo {author} {\bibfnamefont {P.~A.}\ \bibnamefont
  {Denissenkov}}, \bibinfo {author} {\bibfnamefont {D.~A.}\ \bibnamefont
  {VandenBerg}}, \bibinfo {author} {\bibfnamefont {F.~D.~A.}\ \bibnamefont
  {Hartwick}}, \bibinfo {author} {\bibfnamefont {F.}~\bibnamefont {Herwig}},
  \bibinfo {author} {\bibfnamefont {A.}~\bibnamefont {Weiss}}, \ and\ \bibinfo
  {author} {\bibfnamefont {B.}~\bibnamefont {Paxton}},\ }\href {\doibase
  10.1093/mnras/stv211} {\bibfield  {journal} {\bibinfo  {journal} {Monthly
  Notices of the Royal Astronomical Society}\ }\textbf {\bibinfo {volume}
  {448}},\ \bibinfo {pages} {3314} (\bibinfo {year} {2015})}\BibitemShut
  {NoStop}%
\bibitem [{\citenamefont {{de Mink}}\ \emph {et~al.}(2009)\citenamefont {{de
  Mink}}, \citenamefont {{Pols}}, \citenamefont {{Langer}},\ and\ \citenamefont
  {{Izzard}}}]{deMink09-AA}%
  \BibitemOpen
  \bibfield  {author} {\bibinfo {author} {\bibfnamefont {S.~E.}\ \bibnamefont
  {{de Mink}}}, \bibinfo {author} {\bibfnamefont {O.~R.}\ \bibnamefont
  {{Pols}}}, \bibinfo {author} {\bibfnamefont {N.}~\bibnamefont {{Langer}}}, \
  and\ \bibinfo {author} {\bibfnamefont {R.~G.}\ \bibnamefont {{Izzard}}},\
  }\href {\doibase 10.1051/0004-6361/200913205} {\bibfield  {journal} {\bibinfo
   {journal} {Astron.~Astrophys.}\ }\textbf {\bibinfo {volume} {507}},\
  \bibinfo {pages} {L1} (\bibinfo {year} {2009})}\BibitemShut {NoStop}%
\bibitem [{\citenamefont {Sills}\ and\ \citenamefont
  {Glebbeek}(2010)}]{Sills10-MNRAS}%
  \BibitemOpen
  \bibfield  {author} {\bibinfo {author} {\bibfnamefont {A.}~\bibnamefont
  {Sills}}\ and\ \bibinfo {author} {\bibfnamefont {E.}~\bibnamefont
  {Glebbeek}},\ }\href {\doibase 10.1111/j.1365-2966.2010.16876.x} {\bibfield
  {journal} {\bibinfo  {journal} {Monthly Notices of the Royal Astronomical
  Society}\ }\textbf {\bibinfo {volume} {407}},\ \bibinfo {pages} {277}
  (\bibinfo {year} {2010})}\BibitemShut {NoStop}%
\bibitem [{\citenamefont {Maccarone}\ and\ \citenamefont
  {Zurek}(2012)}]{Maccarone12-MNRAS}%
  \BibitemOpen
  \bibfield  {author} {\bibinfo {author} {\bibfnamefont {T.~J.}\ \bibnamefont
  {Maccarone}}\ and\ \bibinfo {author} {\bibfnamefont {D.~R.}\ \bibnamefont
  {Zurek}},\ }\href {\doibase 10.1111/j.1365-2966.2011.20328.x} {\bibfield
  {journal} {\bibinfo  {journal} {Monthly Notices of the Royal Astronomical
  Society}\ }\textbf {\bibinfo {volume} {423}},\ \bibinfo {pages} {2} (\bibinfo
  {year} {2012})}\BibitemShut {NoStop}%
\bibitem [{\citenamefont {{Angulo}}\ \emph {et~al.}(1999)\citenamefont
  {{Angulo}}, \citenamefont {{Arnould}}, \citenamefont {{Rayet}}, \citenamefont
  {{Descouvemont}}, \citenamefont {{Baye}}, \citenamefont {{Leclercq-Willain}},
  \citenamefont {{Coc}}, \citenamefont {{Barhoumi}}, \citenamefont {{Aguer}},
  \citenamefont {{Rolfs}}, \citenamefont {{Kunz}}, \citenamefont {{Hammer}},
  \citenamefont {{Mayer}}, \citenamefont {{Paradellis}}, \citenamefont
  {{Kossionides}}, \citenamefont {{Chronidou}}, \citenamefont {{Spyrou}},
  \citenamefont {{degl'Innocenti}}, \citenamefont {{Fiorentini}}, \citenamefont
  {{Ricci}}, \citenamefont {{Zavatarelli}}, \citenamefont {{Providencia}},
  \citenamefont {{Wolters}}, \citenamefont {{Soares}}, \citenamefont {{Grama}},
  \citenamefont {{Rahighi}}, \citenamefont {{Shotter}},\ and\ \citenamefont
  {{Lamehi Rachti}}}]{NACRE99-NPA}%
  \BibitemOpen
  \bibfield  {author} {\bibinfo {author} {\bibfnamefont {C.}~\bibnamefont
  {{Angulo}}}, \bibinfo {author} {\bibfnamefont {M.}~\bibnamefont {{Arnould}}},
  \bibinfo {author} {\bibfnamefont {M.}~\bibnamefont {{Rayet}}}, \bibinfo
  {author} {\bibfnamefont {P.}~\bibnamefont {{Descouvemont}}}, \bibinfo
  {author} {\bibfnamefont {D.}~\bibnamefont {{Baye}}}, \bibinfo {author}
  {\bibfnamefont {C.}~\bibnamefont {{Leclercq-Willain}}}, \bibinfo {author}
  {\bibfnamefont {A.}~\bibnamefont {{Coc}}}, \bibinfo {author} {\bibfnamefont
  {S.}~\bibnamefont {{Barhoumi}}}, \bibinfo {author} {\bibfnamefont
  {P.}~\bibnamefont {{Aguer}}}, \bibinfo {author} {\bibfnamefont
  {C.}~\bibnamefont {{Rolfs}}}, \bibinfo {author} {\bibfnamefont
  {R.}~\bibnamefont {{Kunz}}}, \bibinfo {author} {\bibfnamefont {J.~W.}\
  \bibnamefont {{Hammer}}}, \bibinfo {author} {\bibfnamefont {A.}~\bibnamefont
  {{Mayer}}}, \bibinfo {author} {\bibfnamefont {T.}~\bibnamefont
  {{Paradellis}}}, \bibinfo {author} {\bibfnamefont {S.}~\bibnamefont
  {{Kossionides}}}, \bibinfo {author} {\bibfnamefont {C.}~\bibnamefont
  {{Chronidou}}}, \bibinfo {author} {\bibfnamefont {K.}~\bibnamefont
  {{Spyrou}}}, \bibinfo {author} {\bibfnamefont {S.}~\bibnamefont
  {{degl'Innocenti}}}, \bibinfo {author} {\bibfnamefont {G.}~\bibnamefont
  {{Fiorentini}}}, \bibinfo {author} {\bibfnamefont {B.}~\bibnamefont
  {{Ricci}}}, \bibinfo {author} {\bibfnamefont {S.}~\bibnamefont
  {{Zavatarelli}}}, \bibinfo {author} {\bibfnamefont {C.}~\bibnamefont
  {{Providencia}}}, \bibinfo {author} {\bibfnamefont {H.}~\bibnamefont
  {{Wolters}}}, \bibinfo {author} {\bibfnamefont {J.}~\bibnamefont {{Soares}}},
  \bibinfo {author} {\bibfnamefont {C.}~\bibnamefont {{Grama}}}, \bibinfo
  {author} {\bibfnamefont {J.}~\bibnamefont {{Rahighi}}}, \bibinfo {author}
  {\bibfnamefont {A.}~\bibnamefont {{Shotter}}}, \ and\ \bibinfo {author}
  {\bibfnamefont {M.}~\bibnamefont {{Lamehi Rachti}}},\ }\href {\doibase
  10.1016/S0375-9474(99)00030-5} {\bibfield  {journal} {\bibinfo  {journal}
  {Nucl.~Phys.~A}\ }\textbf {\bibinfo {volume} {656}},\ \bibinfo {pages} {3}
  (\bibinfo {year} {1999})}\BibitemShut {NoStop}%
\bibitem [{\citenamefont {{Iliadis}}\ \emph
  {et~al.}(2010{\natexlab{a}})\citenamefont {{Iliadis}}, \citenamefont
  {{Longland}}, \citenamefont {{Champagne}}, \citenamefont {{Coc}},\ and\
  \citenamefont {{Fitzgerald}}}]{Iliadis10-NPA841_31}%
  \BibitemOpen
  \bibfield  {author} {\bibinfo {author} {\bibfnamefont {C.}~\bibnamefont
  {{Iliadis}}}, \bibinfo {author} {\bibfnamefont {R.}~\bibnamefont
  {{Longland}}}, \bibinfo {author} {\bibfnamefont {A.~E.}\ \bibnamefont
  {{Champagne}}}, \bibinfo {author} {\bibfnamefont {A.}~\bibnamefont {{Coc}}},
  \ and\ \bibinfo {author} {\bibfnamefont {R.}~\bibnamefont {{Fitzgerald}}},\
  }\href {\doibase 10.1016/j.nuclphysa.2010.04.009} {\bibfield  {journal}
  {\bibinfo  {journal} {Nucl.~Phys.~A}\ }\textbf {\bibinfo {volume} {841}},\
  \bibinfo {pages} {31} (\bibinfo {year} {2010}{\natexlab{a}})}\BibitemShut
  {NoStop}%
\bibitem [{\citenamefont {{Iliadis}}\ \emph
  {et~al.}(2010{\natexlab{b}})\citenamefont {{Iliadis}}, \citenamefont
  {{Longland}}, \citenamefont {{Champagne}},\ and\ \citenamefont
  {{Coc}}}]{Iliadis10-NPA841_251}%
  \BibitemOpen
  \bibfield  {author} {\bibinfo {author} {\bibfnamefont {C.}~\bibnamefont
  {{Iliadis}}}, \bibinfo {author} {\bibfnamefont {R.}~\bibnamefont
  {{Longland}}}, \bibinfo {author} {\bibfnamefont {A.~E.}\ \bibnamefont
  {{Champagne}}}, \ and\ \bibinfo {author} {\bibfnamefont {A.}~\bibnamefont
  {{Coc}}},\ }\href {\doibase 10.1016/j.nuclphysa.2010.04.010} {\bibfield
  {journal} {\bibinfo  {journal} {Nucl.~Phys.~A}\ }\textbf {\bibinfo {volume}
  {841}},\ \bibinfo {pages} {251} (\bibinfo {year}
  {2010}{\natexlab{b}})}\BibitemShut {NoStop}%
\bibitem [{\citenamefont {{Sallaska}}\ \emph {et~al.}(2013)\citenamefont
  {{Sallaska}}, \citenamefont {{Iliadis}}, \citenamefont {{Champange}},
  \citenamefont {{Goriely}}, \citenamefont {{Starrfield}},\ and\ \citenamefont
  {{Timmes}}}]{Sallaska13-ApJSS}%
  \BibitemOpen
  \bibfield  {author} {\bibinfo {author} {\bibfnamefont {A.~L.}\ \bibnamefont
  {{Sallaska}}}, \bibinfo {author} {\bibfnamefont {C.}~\bibnamefont
  {{Iliadis}}}, \bibinfo {author} {\bibfnamefont {A.~E.}\ \bibnamefont
  {{Champange}}}, \bibinfo {author} {\bibfnamefont {S.}~\bibnamefont
  {{Goriely}}}, \bibinfo {author} {\bibfnamefont {S.}~\bibnamefont
  {{Starrfield}}}, \ and\ \bibinfo {author} {\bibfnamefont {F.~X.}\
  \bibnamefont {{Timmes}}},\ }\href {\doibase 10.1088/0067-0049/207/1/18}
  {\bibfield  {journal} {\bibinfo  {journal} {Astrophys.~J.~Suppl.~Ser.}\
  }\textbf {\bibinfo {volume} {207}},\ \bibinfo {eid} {18} (\bibinfo {year}
  {2013})}\BibitemShut {NoStop}%
\bibitem [{\citenamefont {{Depalo}}\ \emph {et~al.}(2015)\citenamefont
  {{Depalo}}, \citenamefont {{Cavanna}}, \citenamefont {{Ferraro}},
  \citenamefont {{Slemer}}, \citenamefont {{Al-Abdullah}}, \citenamefont
  {{Akhmadaliev}}, \citenamefont {{Anders}}, \citenamefont {{Bemmerer}},
  \citenamefont {{Elekes}}, \citenamefont {{Mattei}}, \citenamefont
  {{Reinicke}}, \citenamefont {{Schmidt}}, \citenamefont {{Scian}},\ and\
  \citenamefont {{Wagner}}}]{Depalo15-PRC}%
  \BibitemOpen
  \bibfield  {author} {\bibinfo {author} {\bibfnamefont {R.}~\bibnamefont
  {{Depalo}}}, \bibinfo {author} {\bibfnamefont {F.}~\bibnamefont {{Cavanna}}},
  \bibinfo {author} {\bibfnamefont {F.}~\bibnamefont {{Ferraro}}}, \bibinfo
  {author} {\bibfnamefont {A.}~\bibnamefont {{Slemer}}}, \bibinfo {author}
  {\bibfnamefont {T.}~\bibnamefont {{Al-Abdullah}}}, \bibinfo {author}
  {\bibfnamefont {S.}~\bibnamefont {{Akhmadaliev}}}, \bibinfo {author}
  {\bibfnamefont {M.}~\bibnamefont {{Anders}}}, \bibinfo {author}
  {\bibfnamefont {D.}~\bibnamefont {{Bemmerer}}}, \bibinfo {author}
  {\bibfnamefont {Z.}~\bibnamefont {{Elekes}}}, \bibinfo {author}
  {\bibfnamefont {G.}~\bibnamefont {{Mattei}}}, \bibinfo {author}
  {\bibfnamefont {S.}~\bibnamefont {{Reinicke}}}, \bibinfo {author}
  {\bibfnamefont {K.}~\bibnamefont {{Schmidt}}}, \bibinfo {author}
  {\bibfnamefont {C.}~\bibnamefont {{Scian}}}, \ and\ \bibinfo {author}
  {\bibfnamefont {L.}~\bibnamefont {{Wagner}}},\ }\href {\doibase
  10.1103/PhysRevC.92.045807} {\bibfield  {journal} {\bibinfo  {journal}
  {Phys.~Rev.~C}\ }\textbf {\bibinfo {volume} {92}},\ \bibinfo {pages} {045807}
  (\bibinfo {year} {2015})}\BibitemShut {NoStop}%
\bibitem [{\citenamefont {{Powers}}\ \emph {et~al.}(1971)\citenamefont
  {{Powers}}, \citenamefont {{Fortune}}, \citenamefont {{Middleton}},\ and\
  \citenamefont {{Hansen}}}]{Powers71-PRC}%
  \BibitemOpen
  \bibfield  {author} {\bibinfo {author} {\bibfnamefont {J.~R.}\ \bibnamefont
  {{Powers}}}, \bibinfo {author} {\bibfnamefont {H.~T.}\ \bibnamefont
  {{Fortune}}}, \bibinfo {author} {\bibfnamefont {R.}~\bibnamefont
  {{Middleton}}}, \ and\ \bibinfo {author} {\bibfnamefont {O.}~\bibnamefont
  {{Hansen}}},\ }\href {\doibase 10.1103/PhysRevC.4.2030} {\bibfield  {journal}
  {\bibinfo  {journal} {Phys.~Rev.~C}\ }\textbf {\bibinfo {volume} {4}},\
  \bibinfo {pages} {2030} (\bibinfo {year} {1971})}\BibitemShut {NoStop}%
\bibitem [{\citenamefont {{Hale}}\ \emph {et~al.}(2001)\citenamefont {{Hale}},
  \citenamefont {{Champagne}}, \citenamefont {{Iliadis}}, \citenamefont
  {{Hansper}}, \citenamefont {{Powell}},\ and\ \citenamefont
  {{Blackmon}}}]{Hale01-PRC}%
  \BibitemOpen
  \bibfield  {author} {\bibinfo {author} {\bibfnamefont {S.~E.}\ \bibnamefont
  {{Hale}}}, \bibinfo {author} {\bibfnamefont {A.~E.}\ \bibnamefont
  {{Champagne}}}, \bibinfo {author} {\bibfnamefont {C.}~\bibnamefont
  {{Iliadis}}}, \bibinfo {author} {\bibfnamefont {V.~Y.}\ \bibnamefont
  {{Hansper}}}, \bibinfo {author} {\bibfnamefont {D.~C.}\ \bibnamefont
  {{Powell}}}, \ and\ \bibinfo {author} {\bibfnamefont {J.~C.}\ \bibnamefont
  {{Blackmon}}},\ }\href {\doibase 10.1103/PhysRevC.65.015801} {\bibfield
  {journal} {\bibinfo  {journal} {Phys.~Rev.~C}\ }\textbf {\bibinfo {volume}
  {65}},\ \bibinfo {pages} {015801} (\bibinfo {year} {2001})}\BibitemShut
  {NoStop}%
\bibitem [{\citenamefont {Jenkins}\ \emph {et~al.}(2013)\citenamefont
  {Jenkins}, \citenamefont {Bouhelal}, \citenamefont {Courtin}, \citenamefont
  {Freer}, \citenamefont {Fulton} \emph {et~al.}}]{Jenkins13-PRC}%
  \BibitemOpen
  \bibfield  {author} {\bibinfo {author} {\bibfnamefont {D.}~\bibnamefont
  {Jenkins}}, \bibinfo {author} {\bibfnamefont {M.}~\bibnamefont {Bouhelal}},
  \bibinfo {author} {\bibfnamefont {S.}~\bibnamefont {Courtin}}, \bibinfo
  {author} {\bibfnamefont {M.}~\bibnamefont {Freer}}, \bibinfo {author}
  {\bibfnamefont {B.}~\bibnamefont {Fulton}},  \emph {et~al.},\ }\href
  {\doibase 10.1103/PhysRevC.87.064301} {\bibfield  {journal} {\bibinfo
  {journal} {Phys.Rev. C}\ }\textbf {\bibinfo {volume} {87}},\ \bibinfo {pages}
  {064301} (\bibinfo {year} {2013})}\BibitemShut {NoStop}%
\bibitem [{\citenamefont {{G\"orres}}\ \emph {et~al.}(1982)\citenamefont
  {{G\"orres}}, \citenamefont {{Rolfs}}, \citenamefont {{Schmalbrock}},
  \citenamefont {{Trautvetter}},\ and\ \citenamefont
  {{Keinonen}}}]{Goerres82-NPA}%
  \BibitemOpen
  \bibfield  {author} {\bibinfo {author} {\bibfnamefont {J.}~\bibnamefont
  {{G\"orres}}}, \bibinfo {author} {\bibfnamefont {C.}~\bibnamefont {{Rolfs}}},
  \bibinfo {author} {\bibfnamefont {P.}~\bibnamefont {{Schmalbrock}}}, \bibinfo
  {author} {\bibfnamefont {H.~P.}\ \bibnamefont {{Trautvetter}}}, \ and\
  \bibinfo {author} {\bibfnamefont {J.}~\bibnamefont {{Keinonen}}},\ }\href
  {\doibase 10.1016/0375-9474(82)90489-4} {\bibfield  {journal} {\bibinfo
  {journal} {Nucl.~Phys.~A}\ }\textbf {\bibinfo {volume} {385}},\ \bibinfo
  {pages} {57} (\bibinfo {year} {1982})}\BibitemShut {NoStop}%
\bibitem [{\citenamefont {{Firestone}}(2007)}]{Firestone07-NDS23}%
  \BibitemOpen
  \bibfield  {author} {\bibinfo {author} {\bibfnamefont {R.~B.}\ \bibnamefont
  {{Firestone}}},\ }\href {\doibase 10.1016/j.nds.2007.01.002} {\bibfield
  {journal} {\bibinfo  {journal} {Nucl.~Data~Sheets}\ }\textbf {\bibinfo
  {volume} {108}},\ \bibinfo {pages} {1} (\bibinfo {year} {2007})}\BibitemShut
  {NoStop}%
\bibitem [{\citenamefont {{Cavanna}}\ \emph {et~al.}(2015)\citenamefont
  {{Cavanna}}, \citenamefont {{Depalo}}, \citenamefont {{Aliotta}},
  \citenamefont {{Anders}}, \citenamefont {{Bemmerer}}, \citenamefont {{Best}},
  \citenamefont {{Boeltzig}}, \citenamefont {{Broggini}}, \citenamefont
  {{Bruno}}, \citenamefont {{Caciolli}}, \citenamefont {{Corvisiero}},
  \citenamefont {{Davinson}}, \citenamefont {{di Leva}}, \citenamefont
  {{Elekes}}, \citenamefont {{Ferraro}}, \citenamefont {{Formicola}},
  \citenamefont {{F{\"u}l{\"o}p}}, \citenamefont {{Gervino}}, \citenamefont
  {{Guglielmetti}}, \citenamefont {{Gustavino}}, \citenamefont {{Gy{\"u}rky}},
  \citenamefont {{Imbriani}}, \citenamefont {{Junker}}, \citenamefont
  {{Menegazzo}}, \citenamefont {{Mossa}}, \citenamefont {{Pantaleo}},
  \citenamefont {{Prati}}, \citenamefont {{Scott}}, \citenamefont {{Somorjai}},
  \citenamefont {{Straniero}}, \citenamefont {{Strieder}}, \citenamefont
  {{Sz{\"u}cs}}, \citenamefont {{Tak{\'a}cs}}, \citenamefont {{Trezzi}},\ and\
  \citenamefont {{LUNA Collaboration}}}]{Cavanna15-PRL}%
  \BibitemOpen
  \bibfield  {author} {\bibinfo {author} {\bibfnamefont {F.}~\bibnamefont
  {{Cavanna}}}, \bibinfo {author} {\bibfnamefont {R.}~\bibnamefont {{Depalo}}},
  \bibinfo {author} {\bibfnamefont {M.}~\bibnamefont {{Aliotta}}}, \bibinfo
  {author} {\bibfnamefont {M.}~\bibnamefont {{Anders}}}, \bibinfo {author}
  {\bibfnamefont {D.}~\bibnamefont {{Bemmerer}}}, \bibinfo {author}
  {\bibfnamefont {A.}~\bibnamefont {{Best}}}, \bibinfo {author} {\bibfnamefont
  {A.}~\bibnamefont {{Boeltzig}}}, \bibinfo {author} {\bibfnamefont
  {C.}~\bibnamefont {{Broggini}}}, \bibinfo {author} {\bibfnamefont {C.~G.}\
  \bibnamefont {{Bruno}}}, \bibinfo {author} {\bibfnamefont {A.}~\bibnamefont
  {{Caciolli}}}, \bibinfo {author} {\bibfnamefont {P.}~\bibnamefont
  {{Corvisiero}}}, \bibinfo {author} {\bibfnamefont {T.}~\bibnamefont
  {{Davinson}}}, \bibinfo {author} {\bibfnamefont {A.}~\bibnamefont {{di
  Leva}}}, \bibinfo {author} {\bibfnamefont {Z.}~\bibnamefont {{Elekes}}},
  \bibinfo {author} {\bibfnamefont {F.}~\bibnamefont {{Ferraro}}}, \bibinfo
  {author} {\bibfnamefont {A.}~\bibnamefont {{Formicola}}}, \bibinfo {author}
  {\bibfnamefont {Z.}~\bibnamefont {{F{\"u}l{\"o}p}}}, \bibinfo {author}
  {\bibfnamefont {G.}~\bibnamefont {{Gervino}}}, \bibinfo {author}
  {\bibfnamefont {A.}~\bibnamefont {{Guglielmetti}}}, \bibinfo {author}
  {\bibfnamefont {C.}~\bibnamefont {{Gustavino}}}, \bibinfo {author}
  {\bibfnamefont {G.}~\bibnamefont {{Gy{\"u}rky}}}, \bibinfo {author}
  {\bibfnamefont {G.}~\bibnamefont {{Imbriani}}}, \bibinfo {author}
  {\bibfnamefont {M.}~\bibnamefont {{Junker}}}, \bibinfo {author}
  {\bibfnamefont {R.}~\bibnamefont {{Menegazzo}}}, \bibinfo {author}
  {\bibfnamefont {V.}~\bibnamefont {{Mossa}}}, \bibinfo {author} {\bibfnamefont
  {F.~R.}\ \bibnamefont {{Pantaleo}}}, \bibinfo {author} {\bibfnamefont
  {P.}~\bibnamefont {{Prati}}}, \bibinfo {author} {\bibfnamefont {D.~A.}\
  \bibnamefont {{Scott}}}, \bibinfo {author} {\bibfnamefont {E.}~\bibnamefont
  {{Somorjai}}}, \bibinfo {author} {\bibfnamefont {O.}~\bibnamefont
  {{Straniero}}}, \bibinfo {author} {\bibfnamefont {F.}~\bibnamefont
  {{Strieder}}}, \bibinfo {author} {\bibfnamefont {T.}~\bibnamefont
  {{Sz{\"u}cs}}}, \bibinfo {author} {\bibfnamefont {M.~P.}\ \bibnamefont
  {{Tak{\'a}cs}}}, \bibinfo {author} {\bibfnamefont {D.}~\bibnamefont
  {{Trezzi}}}, \ and\ \bibinfo {author} {\bibnamefont {{LUNA Collaboration}}},\
  }\href {\doibase 10.1103/PhysRevLett.115.252501} {\bibfield  {journal}
  {\bibinfo  {journal} {Phys.~Rev.~Lett.}\ }\textbf {\bibinfo {volume} {115}},\
  \bibinfo {eid} {252501} (\bibinfo {year} {2015})}\BibitemShut {NoStop}%
\bibitem [{\citenamefont {{Izzard}}\ \emph {et~al.}(2007)\citenamefont
  {{Izzard}}, \citenamefont {{Lugaro}}, \citenamefont {{Karakas}},
  \citenamefont {{Iliadis}},\ and\ \citenamefont {{van Raai}}}]{Izzard07-AA}%
  \BibitemOpen
  \bibfield  {author} {\bibinfo {author} {\bibfnamefont {R.~G.}\ \bibnamefont
  {{Izzard}}}, \bibinfo {author} {\bibfnamefont {M.}~\bibnamefont {{Lugaro}}},
  \bibinfo {author} {\bibfnamefont {A.~I.}\ \bibnamefont {{Karakas}}}, \bibinfo
  {author} {\bibfnamefont {C.}~\bibnamefont {{Iliadis}}}, \ and\ \bibinfo
  {author} {\bibfnamefont {M.}~\bibnamefont {{van Raai}}},\ }\href {\doibase
  10.1051/0004-6361:20066903} {\bibfield  {journal} {\bibinfo  {journal}
  {Astron.~Astrophys.}\ }\textbf {\bibinfo {volume} {466}},\ \bibinfo {pages}
  {641} (\bibinfo {year} {2007})}\BibitemShut {NoStop}%
\bibitem [{\citenamefont {Iliadis}\ \emph {et~al.}(2002)\citenamefont
  {Iliadis}, \citenamefont {Champagne}, \citenamefont {Jos{\'e}}, \citenamefont
  {Starrfield},\ and\ \citenamefont {Tupper}}]{Iliadis02-ApJSS}%
  \BibitemOpen
  \bibfield  {author} {\bibinfo {author} {\bibfnamefont {C.}~\bibnamefont
  {Iliadis}}, \bibinfo {author} {\bibfnamefont {A.}~\bibnamefont {Champagne}},
  \bibinfo {author} {\bibfnamefont {J.}~\bibnamefont {Jos{\'e}}}, \bibinfo
  {author} {\bibfnamefont {S.}~\bibnamefont {Starrfield}}, \ and\ \bibinfo
  {author} {\bibfnamefont {P.}~\bibnamefont {Tupper}},\ }\href@noop {}
  {\bibfield  {journal} {\bibinfo  {journal} {Astrophys.~J.~Suppl.~Ser.}\
  }\textbf {\bibinfo {volume} {142}},\ \bibinfo {pages} {105} (\bibinfo {year}
  {2002})}\BibitemShut {NoStop}%
\bibitem [{\citenamefont {Broggini}\ \emph {et~al.}(2010)\citenamefont
  {Broggini}, \citenamefont {Bemmerer}, \citenamefont {Guglielmetti},\ and\
  \citenamefont {Menegazzo}}]{Broggini10-ARNPS}%
  \BibitemOpen
  \bibfield  {author} {\bibinfo {author} {\bibfnamefont {C.}~\bibnamefont
  {Broggini}}, \bibinfo {author} {\bibfnamefont {D.}~\bibnamefont {Bemmerer}},
  \bibinfo {author} {\bibfnamefont {A.}~\bibnamefont {Guglielmetti}}, \ and\
  \bibinfo {author} {\bibfnamefont {R.}~\bibnamefont {Menegazzo}},\ }\href
  {\doibase 10.1146/annurev.nucl.012809.104526} {\bibfield  {journal} {\bibinfo
   {journal} {Annu. Rev. Nucl. Part. Sci.}\ }\textbf {\bibinfo {volume} {60}},\
  \bibinfo {pages} {53} (\bibinfo {year} {2010})}\BibitemShut {NoStop}%
\bibitem [{\citenamefont {{Formicola}}\ \emph {et~al.}(2003)\citenamefont
  {{Formicola}}, \citenamefont {{Imbriani}}, \citenamefont {{Junker}},
  \citenamefont {{Bemmerer}}, \citenamefont {{Bonetti}}, \citenamefont
  {{Broggini}}, \citenamefont {{Casella}}, \citenamefont {{Corvisiero}},
  \citenamefont {{Costantini}}, \citenamefont {{Gervino}}, \citenamefont
  {{Gustavino}}, \citenamefont {{Lemut}}, \citenamefont {{Prati}},
  \citenamefont {{Roca}}, \citenamefont {{Rolfs}}, \citenamefont {{Romano}},
  \citenamefont {{Sch{\"u}rmann}}, \citenamefont {{Strieder}}, \citenamefont
  {{Terrasi}}, \citenamefont {{Trautvetter}},\ and\ \citenamefont
  {{Zavatarelli}}}]{Formicola03-NIMA}%
  \BibitemOpen
  \bibfield  {author} {\bibinfo {author} {\bibfnamefont {A.}~\bibnamefont
  {{Formicola}}}, \bibinfo {author} {\bibfnamefont {G.}~\bibnamefont
  {{Imbriani}}}, \bibinfo {author} {\bibfnamefont {M.}~\bibnamefont
  {{Junker}}}, \bibinfo {author} {\bibfnamefont {D.}~\bibnamefont
  {{Bemmerer}}}, \bibinfo {author} {\bibfnamefont {R.}~\bibnamefont
  {{Bonetti}}}, \bibinfo {author} {\bibfnamefont {C.}~\bibnamefont
  {{Broggini}}}, \bibinfo {author} {\bibfnamefont {C.}~\bibnamefont
  {{Casella}}}, \bibinfo {author} {\bibfnamefont {P.}~\bibnamefont
  {{Corvisiero}}}, \bibinfo {author} {\bibfnamefont {H.}~\bibnamefont
  {{Costantini}}}, \bibinfo {author} {\bibfnamefont {G.}~\bibnamefont
  {{Gervino}}}, \bibinfo {author} {\bibfnamefont {C.}~\bibnamefont
  {{Gustavino}}}, \bibinfo {author} {\bibfnamefont {A.}~\bibnamefont
  {{Lemut}}}, \bibinfo {author} {\bibfnamefont {P.}~\bibnamefont {{Prati}}},
  \bibinfo {author} {\bibfnamefont {V.}~\bibnamefont {{Roca}}}, \bibinfo
  {author} {\bibfnamefont {C.}~\bibnamefont {{Rolfs}}}, \bibinfo {author}
  {\bibfnamefont {M.}~\bibnamefont {{Romano}}}, \bibinfo {author}
  {\bibfnamefont {D.}~\bibnamefont {{Sch{\"u}rmann}}}, \bibinfo {author}
  {\bibfnamefont {F.}~\bibnamefont {{Strieder}}}, \bibinfo {author}
  {\bibfnamefont {F.}~\bibnamefont {{Terrasi}}}, \bibinfo {author}
  {\bibfnamefont {H.-P.}\ \bibnamefont {{Trautvetter}}}, \ and\ \bibinfo
  {author} {\bibfnamefont {S.}~\bibnamefont {{Zavatarelli}}},\ }\href {\doibase
  10.1016/S0168-9002(03)01435-9} {\bibfield  {journal} {\bibinfo  {journal}
  {Nucl.~Inst.~Meth.~A}\ }\textbf {\bibinfo {volume} {507}},\ \bibinfo {pages}
  {609} (\bibinfo {year} {2003})}\BibitemShut {NoStop}%
\bibitem [{\citenamefont {{Costantini}}\ \emph {et~al.}(2009)\citenamefont
  {{Costantini}}, \citenamefont {Formicola}, \citenamefont {Imbriani},
  \citenamefont {Junker}, \citenamefont {Rolfs},\ and\ \citenamefont
  {Strieder}}]{Costantini09-RPP}%
  \BibitemOpen
  \bibfield  {author} {\bibinfo {author} {\bibfnamefont {H.}~\bibnamefont
  {{Costantini}}}, \bibinfo {author} {\bibfnamefont {A.}~\bibnamefont
  {Formicola}}, \bibinfo {author} {\bibfnamefont {G.}~\bibnamefont {Imbriani}},
  \bibinfo {author} {\bibfnamefont {M.}~\bibnamefont {Junker}}, \bibinfo
  {author} {\bibfnamefont {C.}~\bibnamefont {Rolfs}}, \ and\ \bibinfo {author}
  {\bibfnamefont {F.}~\bibnamefont {Strieder}},\ }\href {\doibase
  10.1088/0034-4885/72/8/086301} {\bibfield  {journal} {\bibinfo  {journal}
  {Rep.~Prog.~Phys.}\ }\textbf {\bibinfo {volume} {72}},\ \bibinfo {pages}
  {086301} (\bibinfo {year} {2009})}\BibitemShut {NoStop}%
\bibitem [{\citenamefont {{Cavanna}}\ \emph {et~al.}(2014)\citenamefont
  {{Cavanna}}, \citenamefont {{Depalo}}, \citenamefont {{Menzel}},
  \citenamefont {{Aliotta}}, \citenamefont {{Anders}}, \citenamefont
  {{Bemmerer}}, \citenamefont {{Broggini}}, \citenamefont {{Bruno}},
  \citenamefont {{Caciolli}}, \citenamefont {{Corvisiero}}, \citenamefont
  {{Davinson}}, \citenamefont {{di Leva}}, \citenamefont {{Elekes}},
  \citenamefont {{Ferraro}}, \citenamefont {{Formicola}}, \citenamefont
  {{F{\"u}l{\"o}p}}, \citenamefont {{Gervino}}, \citenamefont {{Guglielmetti}},
  \citenamefont {{Gustavino}}, \citenamefont {{Gy{\"u}rky}}, \citenamefont
  {{Imbriani}}, \citenamefont {{Junker}}, \citenamefont {{Menegazzo}},
  \citenamefont {{Prati}}, \citenamefont {{Rossi Alvarez}}, \citenamefont
  {{Scott}}, \citenamefont {{Somorjai}}, \citenamefont {{Straniero}},
  \citenamefont {{Strieder}}, \citenamefont {{Sz{\"u}cs}},\ and\ \citenamefont
  {{Trezzi}}}]{Cavanna14-EPJA}%
  \BibitemOpen
  \bibfield  {author} {\bibinfo {author} {\bibfnamefont {F.}~\bibnamefont
  {{Cavanna}}}, \bibinfo {author} {\bibfnamefont {R.}~\bibnamefont {{Depalo}}},
  \bibinfo {author} {\bibfnamefont {M.-L.}\ \bibnamefont {{Menzel}}}, \bibinfo
  {author} {\bibfnamefont {M.}~\bibnamefont {{Aliotta}}}, \bibinfo {author}
  {\bibfnamefont {M.}~\bibnamefont {{Anders}}}, \bibinfo {author}
  {\bibfnamefont {D.}~\bibnamefont {{Bemmerer}}}, \bibinfo {author}
  {\bibfnamefont {C.}~\bibnamefont {{Broggini}}}, \bibinfo {author}
  {\bibfnamefont {C.~G.}\ \bibnamefont {{Bruno}}}, \bibinfo {author}
  {\bibfnamefont {A.}~\bibnamefont {{Caciolli}}}, \bibinfo {author}
  {\bibfnamefont {P.}~\bibnamefont {{Corvisiero}}}, \bibinfo {author}
  {\bibfnamefont {T.}~\bibnamefont {{Davinson}}}, \bibinfo {author}
  {\bibfnamefont {A.}~\bibnamefont {{di Leva}}}, \bibinfo {author}
  {\bibfnamefont {Z.}~\bibnamefont {{Elekes}}}, \bibinfo {author}
  {\bibfnamefont {F.}~\bibnamefont {{Ferraro}}}, \bibinfo {author}
  {\bibfnamefont {A.}~\bibnamefont {{Formicola}}}, \bibinfo {author}
  {\bibfnamefont {Z.}~\bibnamefont {{F{\"u}l{\"o}p}}}, \bibinfo {author}
  {\bibfnamefont {G.}~\bibnamefont {{Gervino}}}, \bibinfo {author}
  {\bibfnamefont {A.}~\bibnamefont {{Guglielmetti}}}, \bibinfo {author}
  {\bibfnamefont {C.}~\bibnamefont {{Gustavino}}}, \bibinfo {author}
  {\bibfnamefont {G.}~\bibnamefont {{Gy{\"u}rky}}}, \bibinfo {author}
  {\bibfnamefont {G.}~\bibnamefont {{Imbriani}}}, \bibinfo {author}
  {\bibfnamefont {M.}~\bibnamefont {{Junker}}}, \bibinfo {author}
  {\bibfnamefont {R.}~\bibnamefont {{Menegazzo}}}, \bibinfo {author}
  {\bibfnamefont {P.}~\bibnamefont {{Prati}}}, \bibinfo {author} {\bibfnamefont
  {C.}~\bibnamefont {{Rossi Alvarez}}}, \bibinfo {author} {\bibfnamefont
  {D.~A.}\ \bibnamefont {{Scott}}}, \bibinfo {author} {\bibfnamefont
  {E.}~\bibnamefont {{Somorjai}}}, \bibinfo {author} {\bibfnamefont
  {O.}~\bibnamefont {{Straniero}}}, \bibinfo {author} {\bibfnamefont
  {F.}~\bibnamefont {{Strieder}}}, \bibinfo {author} {\bibfnamefont
  {T.}~\bibnamefont {{Sz{\"u}cs}}}, \ and\ \bibinfo {author} {\bibfnamefont
  {D.}~\bibnamefont {{Trezzi}}},\ }\href {\doibase 10.1140/epja/i2014-14179-5}
  {\bibfield  {journal} {\bibinfo  {journal} {Eur.~Phys.~J.~A}\ }\textbf
  {\bibinfo {volume} {50}},\ \bibinfo {pages} {179} (\bibinfo {year}
  {2014})}\BibitemShut {NoStop}%
\bibitem [{\citenamefont {{Casella}}\ \emph {et~al.}(2002)\citenamefont
  {{Casella}}, \citenamefont {{Costantini}}, \citenamefont {{Lemut}},
  \citenamefont {{Limata}}, \citenamefont {{Bemmerer}}, \citenamefont
  {{Bonetti}}, \citenamefont {{Broggini}}, \citenamefont {{Campajola}},
  \citenamefont {{Cocconi}}, \citenamefont {{Corvisiero}}, \citenamefont
  {{Cruz}}, \citenamefont {{D'Onofrio}}, \citenamefont {{Formicola}},
  \citenamefont {{F{\"u}l{\"o}p}}, \citenamefont {{Gervino}}, \citenamefont
  {{Gialanella}}, \citenamefont {{Guglielmetti}}, \citenamefont {{Gustavino}},
  \citenamefont {{Gyurky}}, \citenamefont {{Loiano}}, \citenamefont
  {{Imbriani}}, \citenamefont {{Jesus}}, \citenamefont {{Junker}},
  \citenamefont {{Musico}}, \citenamefont {{Ordine}}, \citenamefont {{Parodi}},
  \citenamefont {{Parolin}}, \citenamefont {{Pinto}}, \citenamefont {{Prati}},
  \citenamefont {{Ribeiro}}, \citenamefont {{Roca}}, \citenamefont {{Rogalla}},
  \citenamefont {{Rolfs}}, \citenamefont {{Romano}}, \citenamefont
  {{Rossi-Alvarez}}, \citenamefont {{Rottura}}, \citenamefont {{Schuemann}},
  \citenamefont {{Somorjai}}, \citenamefont {{Strieder}}, \citenamefont
  {{Terrasi}}, \citenamefont {{Trautvetter}}, \citenamefont {{Vomiero}},\ and\
  \citenamefont {{Zavatarelli}}}]{Casella02-NIMA}%
  \BibitemOpen
  \bibfield  {author} {\bibinfo {author} {\bibfnamefont {C.}~\bibnamefont
  {{Casella}}}, \bibinfo {author} {\bibfnamefont {H.}~\bibnamefont
  {{Costantini}}}, \bibinfo {author} {\bibfnamefont {A.}~\bibnamefont
  {{Lemut}}}, \bibinfo {author} {\bibfnamefont {B.}~\bibnamefont {{Limata}}},
  \bibinfo {author} {\bibfnamefont {D.}~\bibnamefont {{Bemmerer}}}, \bibinfo
  {author} {\bibfnamefont {R.}~\bibnamefont {{Bonetti}}}, \bibinfo {author}
  {\bibfnamefont {C.}~\bibnamefont {{Broggini}}}, \bibinfo {author}
  {\bibfnamefont {L.}~\bibnamefont {{Campajola}}}, \bibinfo {author}
  {\bibfnamefont {P.}~\bibnamefont {{Cocconi}}}, \bibinfo {author}
  {\bibfnamefont {P.}~\bibnamefont {{Corvisiero}}}, \bibinfo {author}
  {\bibfnamefont {J.}~\bibnamefont {{Cruz}}}, \bibinfo {author} {\bibfnamefont
  {A.}~\bibnamefont {{D'Onofrio}}}, \bibinfo {author} {\bibfnamefont
  {A.}~\bibnamefont {{Formicola}}}, \bibinfo {author} {\bibfnamefont
  {Z.}~\bibnamefont {{F{\"u}l{\"o}p}}}, \bibinfo {author} {\bibfnamefont
  {G.}~\bibnamefont {{Gervino}}}, \bibinfo {author} {\bibfnamefont
  {L.}~\bibnamefont {{Gialanella}}}, \bibinfo {author} {\bibfnamefont
  {A.}~\bibnamefont {{Guglielmetti}}}, \bibinfo {author} {\bibfnamefont
  {C.}~\bibnamefont {{Gustavino}}}, \bibinfo {author} {\bibfnamefont
  {G.}~\bibnamefont {{Gyurky}}}, \bibinfo {author} {\bibfnamefont
  {A.}~\bibnamefont {{Loiano}}}, \bibinfo {author} {\bibfnamefont
  {G.}~\bibnamefont {{Imbriani}}}, \bibinfo {author} {\bibfnamefont {A.~P.}\
  \bibnamefont {{Jesus}}}, \bibinfo {author} {\bibfnamefont {M.}~\bibnamefont
  {{Junker}}}, \bibinfo {author} {\bibfnamefont {P.}~\bibnamefont {{Musico}}},
  \bibinfo {author} {\bibfnamefont {A.}~\bibnamefont {{Ordine}}}, \bibinfo
  {author} {\bibfnamefont {F.}~\bibnamefont {{Parodi}}}, \bibinfo {author}
  {\bibfnamefont {M.}~\bibnamefont {{Parolin}}}, \bibinfo {author}
  {\bibfnamefont {J.~V.}\ \bibnamefont {{Pinto}}}, \bibinfo {author}
  {\bibfnamefont {P.}~\bibnamefont {{Prati}}}, \bibinfo {author} {\bibfnamefont
  {J.~P.}\ \bibnamefont {{Ribeiro}}}, \bibinfo {author} {\bibfnamefont
  {V.}~\bibnamefont {{Roca}}}, \bibinfo {author} {\bibfnamefont
  {D.}~\bibnamefont {{Rogalla}}}, \bibinfo {author} {\bibfnamefont
  {C.}~\bibnamefont {{Rolfs}}}, \bibinfo {author} {\bibfnamefont
  {M.}~\bibnamefont {{Romano}}}, \bibinfo {author} {\bibfnamefont
  {C.}~\bibnamefont {{Rossi-Alvarez}}}, \bibinfo {author} {\bibfnamefont
  {A.}~\bibnamefont {{Rottura}}}, \bibinfo {author} {\bibfnamefont
  {F.}~\bibnamefont {{Schuemann}}}, \bibinfo {author} {\bibfnamefont
  {E.}~\bibnamefont {{Somorjai}}}, \bibinfo {author} {\bibfnamefont
  {F.}~\bibnamefont {{Strieder}}}, \bibinfo {author} {\bibfnamefont
  {F.}~\bibnamefont {{Terrasi}}}, \bibinfo {author} {\bibfnamefont {H.~P.}\
  \bibnamefont {{Trautvetter}}}, \bibinfo {author} {\bibfnamefont
  {A.}~\bibnamefont {{Vomiero}}}, \ and\ \bibinfo {author} {\bibfnamefont
  {S.}~\bibnamefont {{Zavatarelli}}},\ }\href {\doibase
  10.1016/S0168-9002(02)00577-6} {\bibfield  {journal} {\bibinfo  {journal}
  {Nucl.~Inst.~Meth.~A}\ }\textbf {\bibinfo {volume} {489}},\ \bibinfo {pages}
  {160} (\bibinfo {year} {2002})}\BibitemShut {NoStop}%
\bibitem [{\citenamefont {Gilmore}(2008)}]{Gilmore08-Book}%
  \BibitemOpen
  \bibfield  {author} {\bibinfo {author} {\bibfnamefont {G.}~\bibnamefont
  {Gilmore}},\ }\href@noop {} {\emph {\bibinfo {title} {{{Practical
  $\gamma$-ray spectrometry, 2nd edition}}}}}\ (\bibinfo  {publisher} {John
  Wiley and Sons},\ \bibinfo {address} {New York},\ \bibinfo {year}
  {2008})\BibitemShut {NoStop}%
\bibitem [{\citenamefont {{Formicola}}\ \emph {et~al.}(2004)\citenamefont
  {{Formicola}}, \citenamefont {{Imbriani}}, \citenamefont {{Costantini}},
  \citenamefont {{Angulo}}, \citenamefont {{Bemmerer}}, \citenamefont
  {{Bonetti}}, \citenamefont {{Broggini}}, \citenamefont {{Corvisiero}},
  \citenamefont {{Cruz}}, \citenamefont {{Descouvemont}}, \citenamefont
  {{F{\"u}l{\"o}p}}, \citenamefont {{Gervino}}, \citenamefont {{Guglielmetti}},
  \citenamefont {{Gustavino}}, \citenamefont {{Gy{\"u}rky}}, \citenamefont
  {{Jesus}}, \citenamefont {{Junker}}, \citenamefont {{Lemut}}, \citenamefont
  {{Menegazzo}}, \citenamefont {{Prati}}, \citenamefont {{Roca}}, \citenamefont
  {{Rolfs}}, \citenamefont {{Romano}}, \citenamefont {{Rossi Alvarez}},
  \citenamefont {{Sch{\"u}mann}}, \citenamefont {{Somorjai}}, \citenamefont
  {{Straniero}}, \citenamefont {{Strieder}}, \citenamefont {{Terrasi}},
  \citenamefont {{Trautvetter}}, \citenamefont {{Vomiero}},\ and\ \citenamefont
  {{Zavatarelli}}}]{Formicola04-PLB}%
  \BibitemOpen
  \bibfield  {author} {\bibinfo {author} {\bibfnamefont {A.}~\bibnamefont
  {{Formicola}}}, \bibinfo {author} {\bibfnamefont {G.}~\bibnamefont
  {{Imbriani}}}, \bibinfo {author} {\bibfnamefont {H.}~\bibnamefont
  {{Costantini}}}, \bibinfo {author} {\bibfnamefont {C.}~\bibnamefont
  {{Angulo}}}, \bibinfo {author} {\bibfnamefont {D.}~\bibnamefont
  {{Bemmerer}}}, \bibinfo {author} {\bibfnamefont {R.}~\bibnamefont
  {{Bonetti}}}, \bibinfo {author} {\bibfnamefont {C.}~\bibnamefont
  {{Broggini}}}, \bibinfo {author} {\bibfnamefont {P.}~\bibnamefont
  {{Corvisiero}}}, \bibinfo {author} {\bibfnamefont {J.}~\bibnamefont
  {{Cruz}}}, \bibinfo {author} {\bibfnamefont {P.}~\bibnamefont
  {{Descouvemont}}}, \bibinfo {author} {\bibfnamefont {Z.}~\bibnamefont
  {{F{\"u}l{\"o}p}}}, \bibinfo {author} {\bibfnamefont {G.}~\bibnamefont
  {{Gervino}}}, \bibinfo {author} {\bibfnamefont {A.}~\bibnamefont
  {{Guglielmetti}}}, \bibinfo {author} {\bibfnamefont {C.}~\bibnamefont
  {{Gustavino}}}, \bibinfo {author} {\bibfnamefont {G.}~\bibnamefont
  {{Gy{\"u}rky}}}, \bibinfo {author} {\bibfnamefont {A.~P.}\ \bibnamefont
  {{Jesus}}}, \bibinfo {author} {\bibfnamefont {M.}~\bibnamefont {{Junker}}},
  \bibinfo {author} {\bibfnamefont {A.}~\bibnamefont {{Lemut}}}, \bibinfo
  {author} {\bibfnamefont {R.}~\bibnamefont {{Menegazzo}}}, \bibinfo {author}
  {\bibfnamefont {P.}~\bibnamefont {{Prati}}}, \bibinfo {author} {\bibfnamefont
  {V.}~\bibnamefont {{Roca}}}, \bibinfo {author} {\bibfnamefont
  {C.}~\bibnamefont {{Rolfs}}}, \bibinfo {author} {\bibfnamefont
  {M.}~\bibnamefont {{Romano}}}, \bibinfo {author} {\bibfnamefont
  {C.}~\bibnamefont {{Rossi Alvarez}}}, \bibinfo {author} {\bibfnamefont
  {F.}~\bibnamefont {{Sch{\"u}mann}}}, \bibinfo {author} {\bibfnamefont
  {E.}~\bibnamefont {{Somorjai}}}, \bibinfo {author} {\bibfnamefont
  {O.}~\bibnamefont {{Straniero}}}, \bibinfo {author} {\bibfnamefont
  {F.}~\bibnamefont {{Strieder}}}, \bibinfo {author} {\bibfnamefont
  {F.}~\bibnamefont {{Terrasi}}}, \bibinfo {author} {\bibfnamefont {H.~P.}\
  \bibnamefont {{Trautvetter}}}, \bibinfo {author} {\bibfnamefont
  {A.}~\bibnamefont {{Vomiero}}}, \ and\ \bibinfo {author} {\bibfnamefont
  {S.}~\bibnamefont {{Zavatarelli}}},\ }\href {\doibase
  10.1016/j.physletb.2004.03.092} {\bibfield  {journal} {\bibinfo  {journal}
  {Phys.~Lett.~B}\ }\textbf {\bibinfo {volume} {591}},\ \bibinfo {pages} {61}
  (\bibinfo {year} {2004})}\BibitemShut {NoStop}%
\bibitem [{\citenamefont {Agostinelli}\ \emph {et~al.}(2003)\citenamefont
  {Agostinelli} \emph {et~al.}}]{Agostinelli03-NIMA}%
  \BibitemOpen
  \bibfield  {author} {\bibinfo {author} {\bibfnamefont {S.}~\bibnamefont
  {Agostinelli}} \emph {et~al.},\ }\href@noop {} {\bibfield  {journal}
  {\bibinfo  {journal} {Nucl.~Inst.~Meth.~A}\ }\textbf {\bibinfo {volume}
  {506}},\ \bibinfo {pages} {250} (\bibinfo {year} {2003})}\BibitemShut
  {NoStop}%
\bibitem [{\citenamefont {Knoll}(2010)}]{Knoll10-book}%
  \BibitemOpen
  \bibfield  {author} {\bibinfo {author} {\bibfnamefont {G.~F.}\ \bibnamefont
  {Knoll}},\ }\href@noop {} {\emph {\bibinfo {title} {{Radiation Detection and
  Measurement}}}},\ \bibinfo {edition} {{4$^{\rm th}$}}\ ed.\ (\bibinfo
  {publisher} {{John Wiley \& Sons}},\ \bibinfo {address} {New York},\ \bibinfo
  {year} {2010})\BibitemShut {NoStop}%
\bibitem [{\citenamefont {Cavanna}(2015)}]{Cavanna15-PhD}%
  \BibitemOpen
  \bibfield  {author} {\bibinfo {author} {\bibfnamefont {F.}~\bibnamefont
  {Cavanna}},\ }\emph {\bibinfo {title} {{A direct measurement of the {\rm
  $^{22}$Ne(p,$\gamma$)$^{23}$Na} reaction down to the energies of
  astrophysical interest}}},\ \href@noop {} {Ph.D. thesis},\ \bibinfo  {school}
  {{Universit\`a degli studi di Genova}} (\bibinfo {year} {2015})\BibitemShut
  {NoStop}%
\bibitem [{\citenamefont {Depalo}(2015)}]{Depalo15-PhD}%
  \BibitemOpen
  \bibfield  {author} {\bibinfo {author} {\bibfnamefont {R.}~\bibnamefont
  {Depalo}},\ }\emph {\bibinfo {title} {{The neon-sodium cycle: Study of the
  {\rm $^{22}$Ne(p,$\gamma$)$^{23}$Na} reaction at astrophysical energies}}},\
  \href@noop {} {Ph.D. thesis},\ \bibinfo  {school} {{Universit\`a degli studi
  di Padova}} (\bibinfo {year} {2015})\BibitemShut {NoStop}%
\bibitem [{\citenamefont {Ziegler}\ \emph {et~al.}(2010)\citenamefont
  {Ziegler}, \citenamefont {Ziegler},\ and\ \citenamefont
  {Biersack}}]{Ziegler10-NIMB}%
  \BibitemOpen
  \bibfield  {author} {\bibinfo {author} {\bibfnamefont {J.~F.}\ \bibnamefont
  {Ziegler}}, \bibinfo {author} {\bibfnamefont {M.~D.}\ \bibnamefont
  {Ziegler}}, \ and\ \bibinfo {author} {\bibfnamefont {J.~P.}\ \bibnamefont
  {Biersack}},\ }\href {\doibase 10.1016/j.nimb.2010.02.091} {\bibfield
  {journal} {\bibinfo  {journal} {Nucl.~Inst.~Meth.~B}\ }\textbf {\bibinfo
  {volume} {268}},\ \bibinfo {pages} {1818} (\bibinfo {year}
  {2010})}\BibitemShut {NoStop}%
\bibitem [{\citenamefont {Audi}\ \emph {et~al.}(2012)\citenamefont {Audi},
  \citenamefont {Wang}, \citenamefont {Wapstra}, \citenamefont {Kondev},
  \citenamefont {MacCormick}, \citenamefont {Xu},\ and\ \citenamefont
  {Pfeiffer}}]{Audi12-Ame2012}%
  \BibitemOpen
  \bibfield  {author} {\bibinfo {author} {\bibfnamefont {G.}~\bibnamefont
  {Audi}}, \bibinfo {author} {\bibfnamefont {M.}~\bibnamefont {Wang}}, \bibinfo
  {author} {\bibfnamefont {A.}~\bibnamefont {Wapstra}}, \bibinfo {author}
  {\bibfnamefont {F.}~\bibnamefont {Kondev}}, \bibinfo {author} {\bibfnamefont
  {M.}~\bibnamefont {MacCormick}}, \bibinfo {author} {\bibfnamefont
  {X.}~\bibnamefont {Xu}}, \ and\ \bibinfo {author} {\bibfnamefont
  {B.}~\bibnamefont {Pfeiffer}},\ }\href
  {http://stacks.iop.org/1674-1137/36/i=12/a=002} {\bibfield  {journal}
  {\bibinfo  {journal} {Chin.~Phys.~C}\ }\textbf {\bibinfo {volume} {36}},\
  \bibinfo {pages} {1287} (\bibinfo {year} {2012})}\BibitemShut {NoStop}%
\bibitem [{\citenamefont {Iliadis}(2007)}]{Iliadis07-Book}%
  \BibitemOpen
  \bibfield  {author} {\bibinfo {author} {\bibfnamefont {C.}~\bibnamefont
  {Iliadis}},\ }\href@noop {} {\emph {\bibinfo {title} {Nuclear Physics of
  Stars}}}\ (\bibinfo  {publisher} {Wiley-VCH},\ \bibinfo {address}
  {Weinheim},\ \bibinfo {year} {2007})\BibitemShut {NoStop}%
\bibitem [{\citenamefont {{G{\"o}rres}}\ \emph {et~al.}(1983)\citenamefont
  {{G{\"o}rres}}, \citenamefont {{Becker}}, \citenamefont {{Buchmann}},
  \citenamefont {{Rolfs}}, \citenamefont {{Schmalbrock}}, \citenamefont
  {{Trautvetter}}, \citenamefont {{Vlieks}}, \citenamefont {{Hammer}},\ and\
  \citenamefont {{Donoghue}}}]{Goerres83-NPA}%
  \BibitemOpen
  \bibfield  {author} {\bibinfo {author} {\bibfnamefont {J.}~\bibnamefont
  {{G{\"o}rres}}}, \bibinfo {author} {\bibfnamefont {H.~W.}\ \bibnamefont
  {{Becker}}}, \bibinfo {author} {\bibfnamefont {L.}~\bibnamefont
  {{Buchmann}}}, \bibinfo {author} {\bibfnamefont {C.}~\bibnamefont {{Rolfs}}},
  \bibinfo {author} {\bibfnamefont {P.}~\bibnamefont {{Schmalbrock}}}, \bibinfo
  {author} {\bibfnamefont {H.~P.}\ \bibnamefont {{Trautvetter}}}, \bibinfo
  {author} {\bibfnamefont {A.}~\bibnamefont {{Vlieks}}}, \bibinfo {author}
  {\bibfnamefont {J.~W.}\ \bibnamefont {{Hammer}}}, \ and\ \bibinfo {author}
  {\bibfnamefont {T.~R.}\ \bibnamefont {{Donoghue}}},\ }\href {\doibase
  10.1016/0375-9474(83)90588-2} {\bibfield  {journal} {\bibinfo  {journal}
  {Nucl.~Phys.~A}\ }\textbf {\bibinfo {volume} {408}},\ \bibinfo {pages} {372}
  (\bibinfo {year} {1983})}\BibitemShut {NoStop}%
\bibitem [{\citenamefont {Rolke}\ \emph {et~al.}(2005)\citenamefont {Rolke},
  \citenamefont {L{\'o}pez},\ and\ \citenamefont {Conrad}}]{Rolke05-NIM}%
  \BibitemOpen
  \bibfield  {author} {\bibinfo {author} {\bibfnamefont {W.~A.}\ \bibnamefont
  {Rolke}}, \bibinfo {author} {\bibfnamefont {A.~M.}\ \bibnamefont
  {L{\'o}pez}}, \ and\ \bibinfo {author} {\bibfnamefont {J.}~\bibnamefont
  {Conrad}},\ }\href {\doibase http://dx.doi.org/10.1016/j.nima.2005.05.068}
  {\bibfield  {journal} {\bibinfo  {journal} {Nuclear Instruments and Methods
  in Physics Research Section A: Accelerators, Spectrometers, Detectors and
  Associated Equipment}\ }\textbf {\bibinfo {volume} {551}},\ \bibinfo {pages}
  {493 } (\bibinfo {year} {2005})}\BibitemShut {NoStop}%
\bibitem [{\citenamefont {Assenbaum}\ \emph {et~al.}(1987)\citenamefont
  {Assenbaum}, \citenamefont {Langanke},\ and\ \citenamefont
  {Rolfs}}]{Assenbaum87-ZPA}%
  \BibitemOpen
  \bibfield  {author} {\bibinfo {author} {\bibfnamefont {H.}~\bibnamefont
  {Assenbaum}}, \bibinfo {author} {\bibfnamefont {K.}~\bibnamefont {Langanke}},
  \ and\ \bibinfo {author} {\bibfnamefont {C.}~\bibnamefont {Rolfs}},\
  }\href@noop {} {\bibfield  {journal} {\bibinfo  {journal} {Z.~Phys.~A}\
  }\textbf {\bibinfo {volume} {327}},\ \bibinfo {pages} {461} (\bibinfo {year}
  {1987})}\BibitemShut {NoStop}%
\bibitem [{\citenamefont {{Huang}}\ \emph {et~al.}(1976)\citenamefont
  {{Huang}}, \citenamefont {{Aoyagi}}, \citenamefont {{Chen}}, \citenamefont
  {{Crasemann}},\ and\ \citenamefont {{Mark}}}]{Huang76-ADNDT}%
  \BibitemOpen
  \bibfield  {author} {\bibinfo {author} {\bibfnamefont {K.-N.}\ \bibnamefont
  {{Huang}}}, \bibinfo {author} {\bibfnamefont {M.}~\bibnamefont {{Aoyagi}}},
  \bibinfo {author} {\bibfnamefont {M.~H.}\ \bibnamefont {{Chen}}}, \bibinfo
  {author} {\bibfnamefont {B.}~\bibnamefont {{Crasemann}}}, \ and\ \bibinfo
  {author} {\bibfnamefont {H.}~\bibnamefont {{Mark}}},\ }\href {\doibase
  10.1016/0092-640X(76)90027-9} {\bibfield  {journal} {\bibinfo  {journal} {At.
  Data Nucl. Data Tables}\ }\textbf {\bibinfo {volume} {18}},\ \bibinfo {pages}
  {243} (\bibinfo {year} {1976})}\BibitemShut {NoStop}%
\bibitem [{\citenamefont {Clayton}(1984)}]{Clayton84-Book}%
  \BibitemOpen
  \bibfield  {author} {\bibinfo {author} {\bibfnamefont {D.~D.}\ \bibnamefont
  {Clayton}},\ }\href@noop {} {\emph {\bibinfo {title} {Principles of Stellar
  Evolution and Nucleosynthesis}}}\ (\bibinfo  {publisher} {University of
  Chicago Press},\ \bibinfo {year} {1984})\BibitemShut {NoStop}%
\bibitem [{\citenamefont {{Longland}}\ \emph {et~al.}(2010)\citenamefont
  {{Longland}}, \citenamefont {{Iliadis}}, \citenamefont {{Champagne}},
  \citenamefont {{Newton}}, \citenamefont {{Ugalde}}, \citenamefont {{Coc}},\
  and\ \citenamefont {{Fitzgerald}}}]{Longland10-NPA841_1}%
  \BibitemOpen
  \bibfield  {author} {\bibinfo {author} {\bibfnamefont {R.}~\bibnamefont
  {{Longland}}}, \bibinfo {author} {\bibfnamefont {C.}~\bibnamefont
  {{Iliadis}}}, \bibinfo {author} {\bibfnamefont {A.~E.}\ \bibnamefont
  {{Champagne}}}, \bibinfo {author} {\bibfnamefont {J.}~\bibnamefont
  {{Newton}}}, \bibinfo {author} {\bibfnamefont {C.}~\bibnamefont {{Ugalde}}},
  \bibinfo {author} {\bibfnamefont {A.}~\bibnamefont {{Coc}}}, \ and\ \bibinfo
  {author} {\bibfnamefont {R.}~\bibnamefont {{Fitzgerald}}},\ }\href {\doibase
  http://dx.doi.org/10.1016/j.nuclphysa.2010.04.008} {\bibfield  {journal}
  {\bibinfo  {journal} {Nucl.~Phys.~A}\ }\textbf {\bibinfo {volume} {841}},\
  \bibinfo {pages} {1 } (\bibinfo {year} {2010})},\ \bibinfo {note} {the 2010
  Evaluation of Monte Carlo based Thermonuclear Reaction Rates}\BibitemShut
  {NoStop}%
\end{thebibliography}
\end{document}